\documentclass[12pt,preprint]{aastex}

\slugcomment{Submitted to Astronomical Journal}
\begin{document}
\title{An HST\ Imaging Survey of Low-Mass Stars in the Chamaeleon I\ Star Forming region }
\author{M. Robberto, L. Spina\altaffilmark{a}}
\affil{Space Telescope Science Institute, Baltimore, MD 21218}
\email{robberto@stsci.edu, lspina@arcetri.astro.it}
\author{N. Da Rio}
\affil{European Space Agency - ESTEC, Keplerlaan 1, 2201 AZ Noordwijk, The Netherlands}
\email{ndario@rssd.esa.int}
\author{D. Apai\altaffilmark{b}} 
\affil{Department of Astronomy, 933 N. Cherry Avenue, Tucson, AZ 85721}
\email{apai@as.arizona.edu}
\author{I. Pascucci}
\affil{Lunar and Planetary Laboratory, 1629 E. University Blvd., Tucson AZ 85721}
\email{pascucci@lpl.arizona.edu}
\author{L. Ricci}
\affil{California Insitute for Technology, MC 249-17, 1200 East California Blvd, Pasadena CA 91125}
\email{lricci@astro.caltech.edu}
\author{C. Goddi, L. Testi}
\affil{European Southern Observatory, Karl Schwarzschild Strasse 2, D-85748 Garching, Germany} 
\email{cgoddi@eso.org, ltesti@eso.org}
\and 
\author{F. Palla, F. Bacciotti}
\affil{\mbox{Osservatorio Astrofisico di Arcetri, Largo E. Fermi 5, I-50125 Firenze, Italy}} 
\email{palla@arcetri.astro.it, fran@arcetri.astro.it}
\altaffiltext{a}{Osservatorio Astrofisico di Arcetri, Largo E. Fermi 5, I-50125 Firenze, Italy}
\altaffiltext{b}{Department of Planetary Sciences, 1629 E. University Blvd., Tucson AZ 85721}
\begin{abstract} 
We present new HST/WFPC2 observations of 20 fields centered around T Tauri stars in the Chamaeleon~I star forming region. Images have been obtained in the F631N ([OI]$\lambda 6300$), F656N (H$\alpha$) and F673N ([SII]$\lambda\lambda6716,6731$) narrow-band filters, plus the Johnson V-band equivalent F547M filter. We detect 31 T Tauri stars falling within our fields. We discuss the optical morphology of 10 sources showing evidence of either binarity, circumstellar material, or mass loss. We supplement our photometry with a compilation of optical, infrared and sub-millimeter data from the literature, together with new sub-mm data for three objects, to build the Spectral Energy Distributions (SED) of 19 single sources. Using an SED model fitting tool, we self-consistently estimate a number of stellar and disk parameters, while mass accretion rates are directly derived from our H$\alpha$ photometry. We find that bolometric luminosities derived from dereddened optical data tend to be underestimated in systems with high $\alpha_{2-24}$ IR spectral index, suggesting that disks seen nearly edge-on may occasionally be interpreted as low luminosity (and therefore more evolved) sources. On the other hand, the same $\alpha_{2-24}$ IR spectral index,  a tracer of the amount of dust in the warmer layers of the circumstellar disks, and the mass accretion rate appear to decay with the isocronal stellar age, suggesting that the observed age spread ($\simeq0.5-5$~Myr) within the cluster is real. Our sample contains a few outliers that may have dissipated their circumstellar disks on shorter time-scale. 
\end{abstract}

\section{Introduction}
The formation of stars, brown dwarfs, and planets is associated with a rich phenomenology spanning almost the entire electromagnetic spectrum. During the star formation process, the circumstellar disk material can be accreted into one (or more) central objects, ejected along the stellar polar axis, dispersed through photo-evaporation or condensed into planetesimals. Each of these processes is traced by a characteristic set of morphological and spectroscopic signatures, such as UV excess for stellar mass accretion \citep{hartmann98, muzerolle01, bouvier07}, shock-excited line-emission from jets and Herbig-Haro objects \citep{bally00, bally07, reipurth03, whelan05, podio06}, CO outflows from the disks \citep{richer00, arce07}, or mid-IR and sub-mm signatures in the spectral energy distribution due to grain growth and possible disk clearing by a forming planetary system \citep[e.g.][]{ricci10, andrews+11, calvet02, dominik07, espaillat08, blum08, pascucci09}. Each tracer provides critical information on different aspects of the protostellar system, but only a comprehensive view allows reconstructing the full evolutionary scenario. In particular, by combining optical and IR observations it is possible to address key open questions like the formation of stellar companions, the fraction of disk mass accreted into the central star vs. time, the timescale of planetesimal formation \citep{krumholz06, meyer06, apai10}.

The Hubble and the Spitzer Space Telescopes offer the best available combination of field-of-view, sensitivity, spatial resolution, and wavelength coverage for studying most of these phenomena \citep{padgett99, odell94, bally00, robberto04, ricci08, allen04, gutermuth04, megeath04, muzerolle04, apai05, pascucci08, luhman08a, luhman08b}. 
The HST has repeatedly targeted the Orion Nebula and its associated young cluster, the archetype of star forming regions \citep{odell94, bally00, robberto04, colgan07, ricci08, odell08}. HST observations resolved about 200 circumstellar disks, tens of jets and provided accurate broad-band photometry needed to determine the fundamental stellar parameters. Unfortunately, the Orion cluster is relatively distant ($\sim$420pc), crowded and projected over the bright M42 HII\ region. Those factors make the Orion Nebula cluster a problematic target for Spitzer, while ground-based mid-IR observations having adequate spatial resolution attain low sensitivity limits longward than $\simeq3.5\mu$m \citep{robberto05, smith05}. 
On the other hand, there are other closer regions that have been extensively investigated by Spitzer, but none of them has been studied with the HST with comparable detail. Among them, the Chamaeleon I region is possibly the best site for combining the unique Spitzer and HST capabilities. It is one of the nearest star-forming regions \citep[$d$=160-170 pc; for a review see][]{luhman08a} and nearly coeval to Orion \citep[age $\sim$2 Myr,] {dario10}, making its low-mass members approximately 10 times brighter than Orion. The cluster is young enough that it retains a significant population of primordial disks, but it is old enough that most of its members are no longer highly obscured by dust (typically $A_V<2$). Like Orion, because of the relatively low extinction, optical wavelengths are accessible for the spectral classification of the stellar population \citep[][and references therein]{comeron04, luhman04} and for measuring accretion diagnostics \citep{mohanty05, muzerolle05}. Optical and near-IR imaging and spectroscopic surveys of Chamaeleon~I have produced an extensive and virtually complete census of the stellar and substellar cluster members in the field \citep{luhman07}: there are 237 known members, 33 of which have spectral types indicative of brown dwarfs ($>M6$). The IMF of Chamaeleon I reaches a maximum at a mass of 0.1-0.15 $M_\sun$, somewhat lower than the IMF peak in Orion.

In this paper we report on a study aimed at probing a sample of Chamaeleon I sources with the Wide Field Planetary Camera 2 (WFPC2) onboard HST. Given the extent of region, about $5.5\degr\times 1.5\degr$, we targeted selected fields centered on brown dwarfs, class I and II Pre-Main-Sequence (PMS) objects measured by Spitzer to uncover substellar companions down to 15 AU separation and to directly image circumstellar disks and jets. We concentrate here on the sample of stellar PMS objects, leaving the discussion of the brown dwarf survey to a second paper (Luhman et al., in preparation). In Section 2 we illustrate our observing and data reduction strategy, detailing the extraction of photometry for both point sources and extended objects. In Section 3 we present the HST photometry, complemented by a compilation of IR and millimeter data available in the literature, including new data in the sub-mm range for 3 sources. We also list the main physical parameters of the stellar sources taken from the literature and derive mass accretion rates from our H$\alpha$ photometry. In Section 4 we illustrate the morphology of individual objects, while in Section 5 we model the SEDs for 19 sources using a SED fitting tool; we derive disk parameters that can be compared wih stellar mass, age, luminosity and mass accretion rate. 
Finally, in Section 6 we summarize our findings.

\section{Observations}
\subsection{HST Data Acquisition and Reduction}
The data presented in this paper have been obtained with the Wide Field Planetary Camera 2 (WFPC2) onboard the Hubble Space Telescope (HST) in early 2009 (HST GO program 11983, P.I. Robberto). These are among the latst data taken with the aging instrument, just preceeding the Servicing Mission 4 which replaced WFPC2 with WFC3. 

We targeted 20 fields centered on T Tauri stars, detecting 18 of them\footnote{The HST archive contains four other fields, centered on 2MASSJ11095493-7635101, HN10E, ISO235 and ISO79. They have been observed but the images appear compromised and have not been used in this work.}. For two sources, ISO-ChaI 150 and Cha J11081938-7731522, known for being highly obscured by circumstellar dust \citep{cambresy98, luhman07}, we can only provide detection upper limits. Other 13 members of the Chamaeleon~I complex lying in our imaged fields have also been identified and are presented in this paper. 
One HST\ orbit was dedicated to each field, placing the target on the standard aperture spot of the WFPC2 Planetary Camera (PC) chip, with 45.5~mas/pixel scale. We did not constrain the telescope roll angle. 

Observations were carried out in the narrow-band filters F631N ([OI] $\lambda6300$) , F656N (H$\alpha$), and F673N (centered on the [SII] $\lambda\lambda6716$, 6731 doublet), plus the F547M\ medium-band filter roughly corresponding to the Johnson V-band. The single exposure times were set at  100~s (F631N), 40~s (F656N), 100~s (F673N) and typically 40~s\ for the F547M filter (only for the brightest stars the F547M\ were shorter to prevent saturation) with four exposures per filter taken in two groups of two. Each group was centered at a slightly different position (``two-point dither") for optimal bad pixels and cosmic ray rejection \citep{biretta08}. 

Each set of four images was processed by the OPUS\ pipeline and combined using the MultiDrizzle software \citep{fruchter09}. We used the MultiDrizzle parameters recommended for two-point dithered observations, treating separately the PC chip from the other 3 Wide Field (WF) chips, due to the different pixel scale.
 
\subsection{Source Identification and Photometry}
The images processed by MultiDrizzle are corrected for geometric distortion introduced by the WFPC2 optics, cleaned from bad pixels and cosmic rays, and recombined into a single integrated image properly oriented in Right Ascension and Declination. We used STARFIND\ (in the STSDAS library of PYRAF) to determine the location on the CCDs of all sources in the field. 

After visually inspecting each individual source to reject false identifications we performed aperture photometry with DAOPHOT using a circular aperture of 3 pixels in radius (corresponding to 0\farcs137 on the PC and 0\farcs 299 on the WF chips). The small extraction radius was chosen to optimally estimate the magnitude of weak sources, the large majority. We did not perform PSF\ photometry due to the non-negligible Charge Transfer Inefficiency trails of the aging instrument. Since the values of the zero points derived from the PHOTOFLAM keyword refer to counts measured within an "infinite" aperture\footnote{in the case of WFPC2 the flux in an infinite aperture is estimated to provide 1.096 times, i.e. a tenth of a magnitude, the counts measured in a 0\farcs5 radius aperture.}, we performed an aperture correction to convert the counts measured in our 3 pixel extraction radius to the value expected in the 0\farcs5 radius associated with the zero points. To this purpose we selected, for each filter and camera, a set of bright unsaturated and isolated sources and compared the results obtained with the two apertures. The average ratio, estimated with a sigma clipping algorithm, provided the aperture correction, together with the corresponding standard deviation. This allowed us to derive absolute magnitudes and errors in the HST\ STMAG system. For the narrowband filters we derived the flux in the more appropriate PHOTFLAM\ (erg cm$^{-2}$s$^{-1}$\AA$^{-1}$) system \citep[for the conversion from counts to flux see][]{baggett02}. Finally, we applied a correction for Charge Transfer Efficiency (CTE) loss following \citet{dolphin09}.

A direct measure of the FWHM reveals that a number of sources are extended. To discriminate between extended and non-extended sources, we used the set of bona-fide point sources to build average Point Spead Functions valid for the PC and WF cameras. The FWHM of these PSFs turned out slightly larger than the theoretical FWHM provided by the HST PSF simulator Tiny Tim (of the order of 2 pixels for both channels), as expected since MultiDrizzle cannot fully recover the optimal PSF with only two pointings. 
The photometry of extended sources was determined using extraction radii large enough to contain all the signal above $\simeq 3\sigma$ sky noise floor. We did not apply any corrections for CTE losses, as this is mostly relevant for small apertures.

\subsection{Sub-mm and mm observations}
As part of a larger project carried our with the Atacama Pathfinder EXperiment (APEX) and with the Australia Telescope Compact Array (ATCA) facilities, we obtained new millimeter wavelength data for three sources: 
ESO H$\alpha$-559 (source \#10) , ISO-ChaI 10 (\#5) and HH48 A (\#7). The first two were observed with APEX, a 12-m submm telescope in Chile's Atacama desert, with the Large APEX BOlometer CAmera \citep[][LABOCA; ]{siringo09} operating at the central frequency of 345 GHz (or 870~$\mu$m). The angular resolution of LABOCA is about 19 arcseconds and its total field of view is about 11 arcminutes. 
The observations, conducted under the project 086.C-0653, were carried out on 2010 October 20-21 for a total of 4.4 h (on source) for ESO H$\alpha$-559 and on 2010 October 29-30 for a total 4.8 h (on source) for ISO-ChaI 10. 
Atmospheric conditions were excellent, with a precipitable water vapor around 0.34 mm, corresponding to a zenith opacity of 0.18. The sky opacity was measured every hour with skydips. The pointing of the telescope was checked every hour on the nearby quasar PKS1057-79.
The absolute flux calibration was performed by observing the secondary calibrator B13134 every one or two hours. The telescope focus was checked by observing the star $\eta$ Carinae, at least once per day.
The observations were performed on the fly mode with a rectangular scanning pattern.
The data were reduced with the BOlometer Array Analysis Software, following the procedures described in \citet{siringo09}.
The total (flux) calibration error is about 20\% and the rms noise level in the final maps was $\sim$5~mJy. We report a clear detection of the submm emission from ESO H$\alpha$-559, but only an upper limit could be derived from our data for ISO-ChaI 10.

HH 48 A was observed at 3.3~mm with ATCA and the new CABB digital filter bank, which provides a total continuum bandwidth of 4 GHz. Observations were carried out at a central frequency of 91.000 GHz (3.294 mm) on 2009 Oct 14. The ATCA array was in the hybrid H168 configuration, providing an angular resolution of about 3 arcsec at 3.3 mm.
The gain was calibrated with frequent observations of 1057-797. The passband was calibrated using 1921-293, and the absolute flux scale was determined through observations of Uranus. The uncertainty on the ATCA calibrated flux is about 30\% at 3.3 mm. The MIRIAD package was used for visibilities calibration, Fourier inversion, deconvolution and imaging. 
The rms noise on the ATCA map was about 0.45 mJy. 

We also searched the literature finding seven more sources previously detected at wavelengths longer than 100~$\mu$m.All  fluxes are reported in Table~\ref{Tab:phot_known}, illustrated in the next section.

\section{Results}
\subsection{Final source list and HST photometry}
\begin{figure}
\centering
\epsscale{1.8}
\plottwo{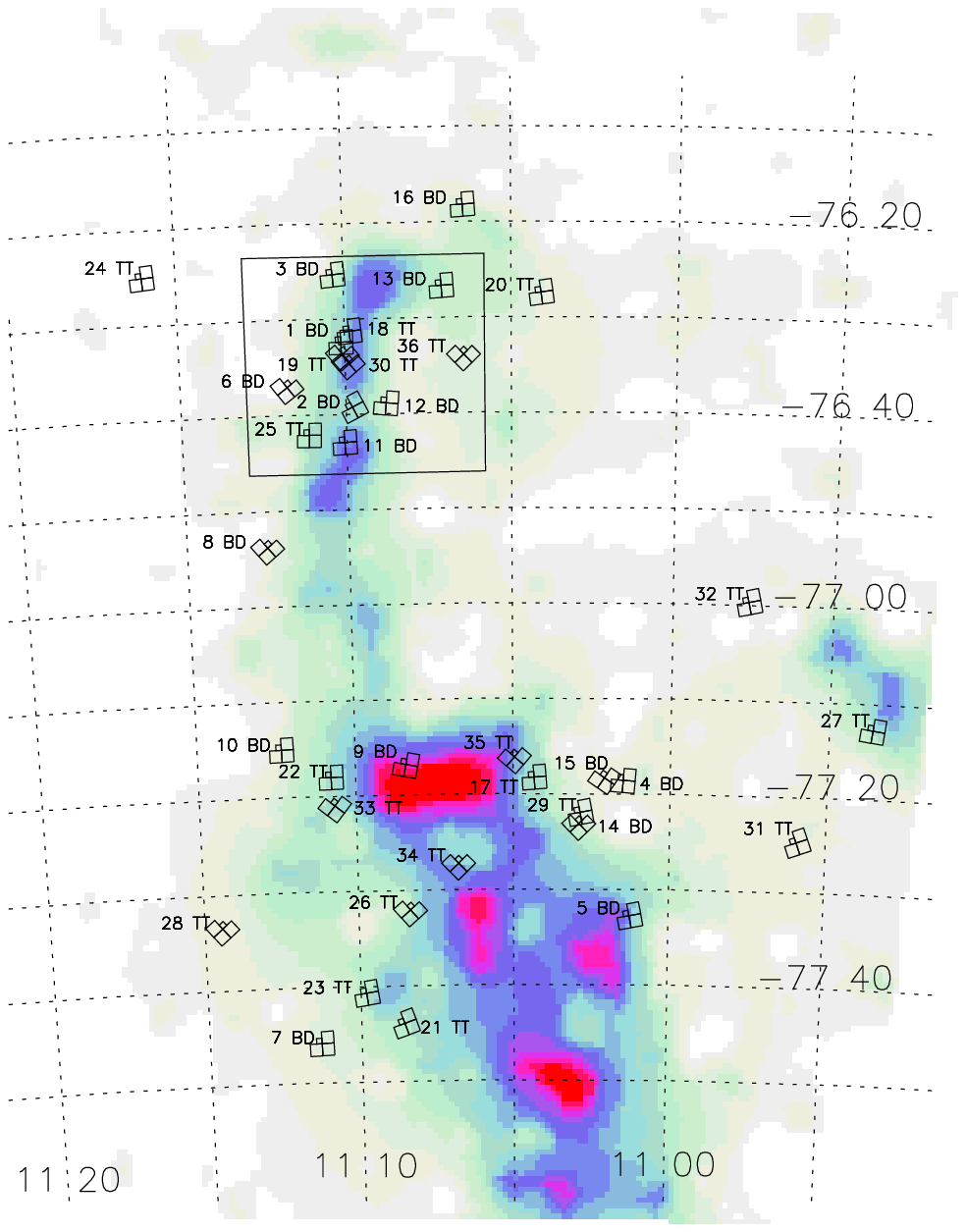}{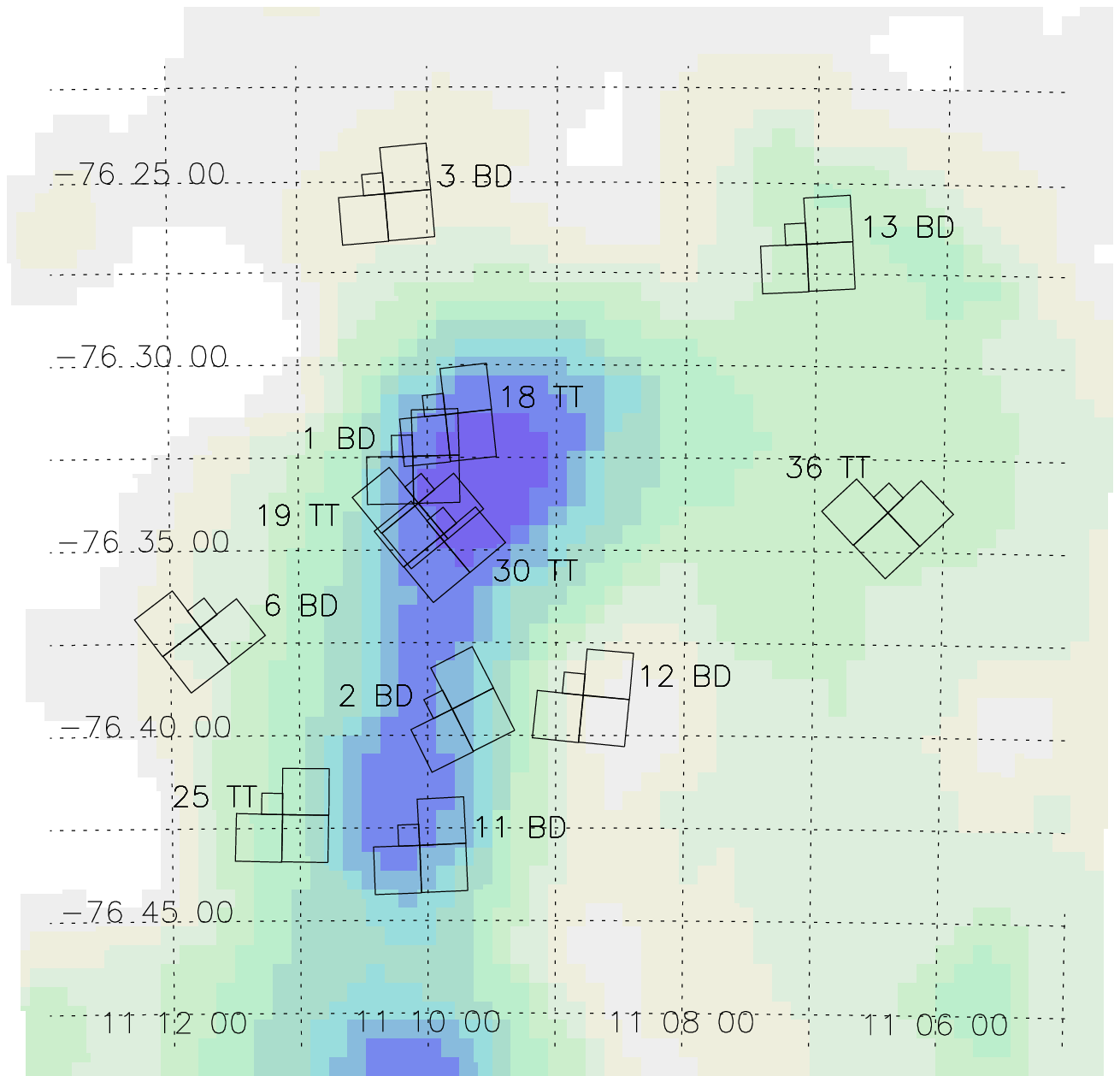}
\caption{All WFPC2 fields observed in our program superimposed to a map of the A$_J$ extinction in Chamaeleon I \citep{cambresy97}. The highest contour corresponds to $A_J=10.7$ while the lowest contour to $A_J=0$. The label of  each field provides the visit number and a suffix, either TT (T Tauri star) or BD (Brown Dwarf), to indicate the nature of the primary target. In this paper we concentrate on the  TT fields. The random orientation of the characteristic WFPC2 chevron pattern is due to unconstrained HST roll angle. For the association between HST visit and the corresponding target see Table 1.\label{field}}
\end{figure}

The location of our 20 WFPC2 fields, each composed of 3 WF images of $\sim72\farcs8\times72\farcs8$ and 1 PC image of $\sim36\farcs4\times36\farcs4$, is shown in Figure \ref{field}. A\ few fields in the northern region appear partially superimposed, but they have been processed and analyzed separately. To facilitate data retrieval, in Table~\ref{Tab:targets} we provide, for each field, the original HST visit number (by order of execution) and target name reported by the HST data archive, together with the more common SIMBAD target name used in this paper. 
\begin{deluxetable}{ccc}
\tablecaption{Main target list\label{Tab:targets}}
\tablehead{\colhead{Visit} & \colhead{HST Target} &\colhead{Source name}}
\startdata
17 & T14A & HH~48~IRS \\ 
18 & ISO225 & ISO-ChaI~225 \\ 
19 & OTS32 & ISO-ChaI~232\\ 
20 & T14 & CT Cha \\ 
21 & IRN & ISO-ChaI 150\tablenotemark{a} \\ 
22 & T47 & HBC 584 \\ 
23 & CHSM15991 & CHSM15991 \\ 
24 & CRHF574 & ESO-H$\alpha$ 574 \\ 
25 & CRHF569 & ESO-H$\alpha$ 569 \\ 
26 & CHAJ11081938-7731522 & Cha J11081938-7731522\tablenotemark{a} \\ 
27 & 2MASSJ10533978-7712338 & 2MASS J10533978-7712338 \\ 
28 & HN21E2 & Hn 21E \\ 
29 & T12 & ISO-ChaI 10 \\ 
30 & T42 & CED 112 IRS 4 \\ 
31 & T3A & SX Cha \\ 
32 & T5 & Ass Cha T 2-5 \\ 
33 & ISO252 & ISO-ChaI 252 \\ 
34 & CHXR20 & UX Cha \\ 
35 & T16 & Ass Cha T 2-16 \\ 
36 & CRHF559& ESO-H$\alpha$ 559 \\ 
\enddata
\tablenotetext{a}{Main target not detected.}
\end{deluxetable}

Besides our 18 detected targets, other 13 objects previously known to be members of the Chamaeleon~I\ association \citep{luhman08b} ended up in our imaged fields. For each of these 31 sources, Table~\ref{Tab:sources} provides: an entry number (column 1), the coordinate-based 2MASS source identifier, when available (column 2); an alternate ID\ taken from the literature
(column 3); equatorial coordinates from the original fits header (column 4 and 5); the HST\ visit, and the (x,y)\ coordinates on the drizzled fits files (column 6 and 7). The accuracy of the celestial coordinates, driven by the absolute positions of the HST guide stars, is typically about $0\farcs25$. 
\begin{deluxetable}{ccccccc}
\rotate 
\tablecaption{Observed members of the Chamaeleon I\ association \label{Tab:sources}} 
\tablewidth{0pt} 
\tablehead{
\colhead{Nr.} &
\colhead{2MASS} & 
\colhead{SIMBAD} & 
\colhead{R.A. (J2000.0)} & 
\colhead{Decl. (J2000.0)} & 
\colhead{Visit} & 
\colhead{FITS COORD. (X-Y)}
}
\startdata
1 & 2MASS J10533978-7712338 & & 10 53 39.78 & -77 12 33.9 				& 27 & 1674-2119 \\
2 & & Ass Cha T 2-3 B & 10 55 59.73 & -77 24 39.9 						& 31 & 1883-2291 \\
3 & & SX Cha & 10 55 59.76 & -77 24 40.1 						& 31 & 1837-2245 \\
4 & 2MASS J10574219-7659356 & Ass Cha T 2-5 & 10 57 42.20 & -76 59 35.7 	& 32 & 1705-2153 \\
5 & 2MASS J11025504-7721508 & ISO-ChaI 10 & 11 02 55.05 & -77 21 50.8 	& 29 & 1681-2143 \\
6 & 2MASS J11040909-7627193 &  CT Cha & 11 04 09.09 & -76 27 19.4 	& 20 & 1631-2084 \\
7 & 2MASS J11042275-7718080 & HH 48 A & 11 04 22.69 & -77 18 09.1 		& 17 & 1545-1991 \\
8 & & HH 48 B & 11 04 23.31 & -77 18 07.5 & 17 & 1502-2024 \\
9 & 2MASS J11045701-7715569 & Ass Cha T 2-16 & 11 04 57.01 & -77 15 56.9 & 35 & 2346-2647 \\
10 & 2MASS J11062554-7633418 & ESO-H$\alpha$ 559 & 11 06 25.55 & -76 33 41.9 & 36 & 2382-2652 \\
11 & 2MASS J11064510-7727023 &  UX Cha & 11 06 45.10 & -77 27 02.3 	& 34 & 2381-2665 \\
12 & 2MASS J11081648-7744371 & Ass Cha T 2-34 & 11 08 16.49 & -77 44 37.2 & 21 & 1551-577 \\
13 & 2MASS J11081703-7744118 & ISO-ChaI 137 & 11 08 17.03 & -77 44 11.8 & 21 & 1534-832 \\
14 & & Cha J11081938-7731522 & 11 08 19.38 & -77 31 52.2 				& 26 & not detected \\
15 & & ISO-ChaI 150 & 11 08 37.1 & -77 43 51 							& 21 & not detected \\
16 & 2MASS J11094525-7740332 & ISO-ChaI 201 & 11 09 45.26 & -77 40 33.3 & 23 & 983-972 \\
17 & 2MASS J11094621-7634463 & Hn 10E & 11 09 46.21 & -76 34 46.4 		& 30 & 1349-996 \\
18 & 2MASS J11095262-7740348 & CHSM 15991 & 11 09 52.62 & -77 40 34.9 	& 23 & 1634-2091 \\
19 & 2MASS J11095340-7634255 & CED 112 IRS 4 & 11 09 53.41 & -76 34 25.5& 30 & 2401-2638 \\
20 & 2MASS J11095437-7631113 & ISO-ChaI 225 & 11 09 54.38 & -76 31 11.4 & 18 & 1585-2054 \\
21 & 2MASS J11100010-7634578 &  WW Cha & 11 10 00.11 & -76 34 57.9 	& 19 & 1212-329 \\
22 & 2MASS J11100369-7633291 & ISO-ChaI 232 & 11 10 03.69 & -76 33 29.2 & 19 & 2386-2684 \\
23 & 2MASS J11100469-7635452 & Cha T 2-45a & 11 10 04.69 & -76 35 45.3 	& 30 & 704-405 \\
24 & 2MASS J11102852-7716596 & Hn 12W & 11 10 28.52 & -77 16 59.6 		& 22 & 1366-1401 \\
25 & 2MASS J11103481-7722053 & [LES2004]\tablenotemark{a} ChaI 405 & 11 10 34.81 & -77 22 05.3 & 33 & 1274-426 \\
26 & 2MASS J11104141-7720480 & ISO-ChaI 252 & 11 10 41.42 & -77 20 48.1 & 33 & 2313-2633 \\
27 & 2MASS J11104959-7717517 & HBC 584 & 11 10 50.00 & -77 17 51.8 	& 22 & 1463-1921 \\
28 & 2MASS J11105076-7718031 & ESO-H$\alpha$ 568 & 11 10 50.77 & -77 18 03.2 & 22 & 630-763 \\
29 & 2MASS J11111083-7641574 & ESO-H$\alpha$ 569 & 11 11 10.83 & -76 41 57.4 & 25 & 1494-1950 \\
30 & 2MASS J11142454-7733062 & Hn 21W & 11 14 24.54 & -77 33 06.2 		& 28 & 2441-2862 \\
31 & 2MASS J11142611-7733042 & Hn 21E & 11 14 26.11 & -77 33 04.3 		& 28 & 2328-2905 \\
32 & 2MASS J11145031-7733390 & BYB 53 & 11 14 50.32 & -77 33 39.0 		& 28 & 278-980 \\
33 & 2MASS J11160287-7624533 & ESO-H$\alpha$ 574 & 11 16 02.88 & -76 24 53.3 & 24 & 1558-2022 \\
\enddata
\tablenotetext{a}{\citet{lopezmarti04}}
\end{deluxetable}

In Table~\ref{Table:TTTab} we provide the WFPC2 photometry together with the photometric extraction area for extended sources. If a source is not detected we report the 3$\sigma$ upper limit estimated over a 3 pixel aperture radius. 

\begin{deluxetable}{cccccc}
\rotate
\tablecaption{Photometry with F547M, F631N, F656N and F673N filters\label{Table:TTTab}}
\tablewidth{0pt}
\tablehead{
\colhead{N.} &
 \colhead{F547M} &
  \colhead{F631N} &
   \colhead{F656N} &
    \colhead{F673N} &
     \colhead{Notes} \\
 & 
 [mag] & 
  [10$^{-16}$ erg s$^{-1}$ cm$^{-2}$ \AA$^{-1}$] & 
   [10$^{-16}$ erg s$^{-1}$ cm$^{-2}$ \AA$^{-1}$] & 
    10$^{-16}$ erg s$^{-1}$ cm$^{-2}$ \AA$^{-1}$] &
}
\startdata
1 & 19.38 $\pm$ 0.18 & $\sim$0.37 & 1.6 $\pm$ 0.9 & 1.1 $\pm$ 0.5 & \\ 
2 & 16.14 $\pm$ 0.08 & 11 $\pm$ 3 & 37 $\pm$ 9 & 16 $\pm$ 4 & \\ 
3 & 16.71 $\pm$ 0.03 & 15 $\pm$ 0.8 & 66 $\pm$ 4 & 15.6 $\pm$ 0.8 & extended; 0\farcs5 aperture \\ 
4 & 14.80 $\pm$ 0.03 & 31.2 $\pm$ 1.7 & 195 $\pm$ 10 & 40 $\pm$ 2 & binary; 0\farcs5 aperture \\ 
5 & 16.25 $\pm$ 0.08 & 10 $\pm$ 3 & 62 $\pm$ 14 & 12 $\pm$ 3 & \\ 
6 & 13.00 $\pm$ 0.07 & $>$230 & $>$770 & $>$250 & companion; 0\farcs5 aperture \\ 
7 & 17.89 $\pm$ 0.03 & 4.1 $\pm$ 0.3 & 17.4 $\pm$ 1.1 & 3.8 $\pm$ 0.2 & disk; 0\farcs5 aperture \\ 
8 & 21.3 $\pm$ 0.3     & 0.32 $\pm$ 0.16 & 2.8 $\pm$ 0.8 & 0.44 $\pm$ 0.11 & disk; 1\arcsec\ aperture \\ 
9 & 18.91 $\pm$ 0.04 & 1.75 $\pm$ 0.15 & 3.4 $\pm$ 0.4 & 2.27 $\pm$ 0.16 & extended; 0\farcs5 aperture \\ 
10 & 20.7 $\pm$ 0.3 & $\sim$0.35 & $\sim$1.3 & $\sim$0.43 & \\ 
11 & 14.78 $\pm$ 0.07 & 88 $\pm$ 18 & 130 $\pm$ 30 & 100 $\pm$ 20 & \\ 
12 & 16.53 $\pm$ 0.05 & 11 $\pm$ 3 & 27 $\pm$ 5 & 15 $\pm$ 3 & \\ 
13 & 18.51 $\pm$ 0.11 & 1.7 $\pm$ 0.7 & 5.3 $\pm$ 1.7 & 2.4 $\pm$ 0.8 & \\ 
14 & $>$24 & $<$0.035 & $<$0.14 & $<$0.023 & \\ 
15 & $>$24 & $<$0.037 & $<$0.15 & $<$0.025 & \\ 
16 & 19.94 $\pm$ 0.02 & 0.62 $\pm$ 0.06 & 2.4 $\pm$ 0.2 & 0.86 $\pm$ 0.06  & \\ 
17 & 18.41 $\pm$ 0.10 & 2.6 $\pm$ 0.9 & 20 $\pm$ 4 & 3.4 $\pm$ 1.0 & \\ 
18 & 21.47 $\pm$ 0.09 & 0.20 $\pm$ 0.04 & 0.32 $\pm$ 0.13 & 0.24 $\pm$ 0.03 & \\ 
19 & 19.50 $\pm$ 0.11 & 2.9 $\pm$ 0.2 & 9.85 $\pm$ 0.7 & 2.15 $\pm$ 0.15 & extended; 0\farcs5 aperture \\ 
20 & 21.6 $\pm$ 0.5 & $\sim$0.13 & $\sim$0.33 & $\sim$0.17 & \\
21 & 14.10 $\pm$ 0.02 & 148 $\pm$ 8 & 650 $\pm$ 30 & 1.49 $\pm$ 8 & extended; 1\arcsec\ aperture \\ 
22 & 18.78 $\pm$ 0.14 & 2.6 $\pm$ 1.0 & 4.6 $\pm$ 1.8 & 4.2 $\pm$ 1.3 & \\ 
23 & 14.68 $\pm$ 0.03 & 70 $\pm$ 12 & 94 $\pm$ 17 & 81 $\pm$ 13 & \\ 
24 & 17.93 $\pm$ 0.09 & 2.8 $\pm$ 1.0 & 8 $\pm$ 2 & 3.5 $\pm$ 1.0 & \\ 
25 & 20.9 $\pm$ 0.3 & $\sim$0.34 & $\sim$1.2 & $\sim$0.52 & \\ 
26 & $\sim$22.4 & $\sim$0.094 & 2.4 $\pm$ 1.2 & $\sim$0.13 & \\ 
27 & 16.64 $\pm$ 0.08 & 10 $\pm$ 3 & 49 $\pm$ 11 & 13 $\pm$ 3 & \\ 
28 & 18.27 $\pm$ 0.03 & 2.72 $\pm$ 0.18 & 6.3 $\pm$ 0.5 & 3.4 $\pm$ 0.2 & \\ 
29 & 20.28 $\pm$ 0.14 & 0.21 $\pm$ 0.15 & 2.4 $\pm$ 0.7 & 0.35 $\pm$ 0.12 & disk; 1\arcsec\ aperture \\ 
30 & 18.20 $\pm$ 0.12 & 2.5 $\pm$ 1.0 & 25 $\pm$ 6 & 3.8 $\pm$ 1.2 & \\ 
31 & 20.7 $\pm$ 0.3 & $\sim$0.28 & $\sim$0.88 & $\sim$0.34 & \\ 
32 & $<$15 & 67 $\pm$ 4 & 99 $\pm$ 5 & 78 $\pm$ 4 & \\ 
33 & 19.60 $\pm$ 0.07 & 1.28 $\pm$ 0.18 & 2.7 $\pm$ 0.6 & 1.96 $\pm$ 0.18 & disk; 1\arcsec\ aperture \\ 
\enddata
\end{deluxetable}

Table~\ref{Tab:phot_known} provides a compilation of near-IR photometric data for the 31 detected sources, either from the ground or from Spitzer \citep{luhman08b, luhmanmuench08}. We also list fluxes at 0.87~mm, 1.3~mm and 3.3~mm from previously published data (\citet{belloche11}, using APEX/LABOCA at 870~micron and $\approx19\arcsec$ angular resolution; \citet{henning93}, using SEST at 1.3~mm and $\approx23\arcsec$ angular resolution), together with our new data for 3 sources (see Section~3.2).
The appropriate reference for each individual source is given in the Appendix. 

\begin{deluxetable}{cccccccccccccc}
\tabletypesize{\tiny}
\rotate
\tablecaption{IR and sub-mm photometry \label{Tab:phot_known}}
\tablewidth{0pt}
\tablehead{
\colhead{N.} & 
\colhead{R\tablenotemark{a}} & 
\colhead{I\tablenotemark{b}} & 
\colhead{J\tablenotemark{c}} & 
\colhead{H\tablenotemark{c}} & 
\colhead{Ks\tablenotemark{c}} &
\colhead{ 3.6 $\mu$m} & 
\colhead{ 4.5 $\mu$m} & 
\colhead{ 5.8 $\mu$m} & 
\colhead{ 8.0 $\mu$m} & 
\colhead{ 24 $\mu$m} & 
\colhead{ $F_{0.87 mm}$} & 
\colhead{ $F_{1.3 mm}$} & 
\colhead{ $F_{3.3 mm}$} \\
& \colhead{SED} \\
& [mag] & [mag] & [mag] & [mag] & [mag] & [mag] & [mag] & [mag] & [mag] & [mag] & [mJy] & [mJy] & [mJy] \\
}
\startdata
1 & & 15.56 $\pm$ 0.06 & 13.28 $\pm$ 0.03 & $12.14\pm 0.02$ &$11.58 \pm 0.02$& 11.49 $\pm$ 0.02 & 11.09 $\pm$ 0.02 & 10.60 $\pm$ 0.03 & 9.76 $\pm$ 0.04 & 5.28 $\pm$ 0.04 & & & \\
2 & & 12.35 $\pm$ 0.02 & & & \\ 
3 & & 12.35 $\pm$ 0.02 & 10.65 $\pm$ 0.02 & 9.84 $\pm$ 0.02 & 8.69 $\pm$ 0.02 \\ 
4 & & 12.17 $\pm$ 0.03 & 10.43 $\pm$ 0.02 & 9.56 $\pm$ 0.02 & 9.25 $\pm$ 0.02 & 8.74 $\pm$ 0.02 & & 8.07 $\pm$ 0.03 & & 4.35 $\pm$ 0.04 & & & \\
5 & 15.17 $\pm$ 0.05 & 13.40 $\pm$ 0.05 & 11.56 $\pm$ 0.03 &10.86 $\pm$ 0.03 & 10.45 $\pm$ 0.02 & 9.78 $\pm$ 0.02 & & 9.20 $\pm$ 0.03 & 8.65 $\pm$ 0.03 & 5.90 $\pm$ 0.04 & $<$14.8 & & \\
6 & & & 9.71 $\pm$ 0.02 & 8.94 $\pm$ 0.05 & 8.66 $\pm$ 0.02 & & & & & 2.44 $\pm$ 0.04 & & & \\
7 & & & 16.42\tablenotemark{4} & 14.32\tablenotemark{4} & 12.54\tablenotemark{4} & 10.06 $\pm$ 0.02 & 9.23 $\pm$ 0.02 & 8.25 $\pm$ 0.03 & 6.97 $\pm$ 0.04 & 3.47 $\pm$ 0.04 & & & 3.95 \\
8 & & & 18.43\tablenotemark{4} & 15.66\tablenotemark{4}  & 13.85\tablenotemark{4} &  &  &  &  &  & & & \\
9 & 16.65 $\pm$ 0.05 & 14.68 $\pm$ 0.04 & 12.17 $\pm$ 0.02 &10.97 $\pm$ 0.02 & 10.41 $\pm$ 0.02 & 9.80 $\pm$ 0.02 & 9.57 $\pm$ 0.02 & 9.18 $\pm$ 0.04 & 8.63 $\pm$ 0.04 & 6.11 $\pm$ 0.04 & & & \\
10 & & 15.88 $\pm$ 0.06 & 13.01 $\pm$ 0.03 &12.01 $\pm$ 0.02 & 11.49 $\pm$ 0.02 & 10.83 $\pm$ 0.02 & 10.42 $\pm$ 0.02 & 10.07 $\pm$ 0.03 & 9.38 $\pm$ 0.04 & 5.34 $\pm$ 0.04 & 44.1 & & \\
11 & &12.07 $\pm$ 0.02 & 10.18 $\pm$ 0.02 & 9.20 $\pm$ 0.02 &8.880 $\pm$ 0.019 & 8.51 $\pm$ 0.02 & 8.36 $\pm$ 0.02 & 7.93 $\pm$ 0.03 & 6.94 $\pm$ 0.04 & 4.45 $\pm$ 0.04 & & & \\
12 & & 13.12 $\pm$ 0.03 & 11.20 $\pm$ 0.03 &10.34 $\pm$ 0.03 & 10.02 $\pm$ 0.02 & 9.78 $\pm$ 0.02 & 9.75 $\pm$ 0.02 & 9.73 $\pm$ 0.03 & 9.71 $\pm$ 0.03 & & & & \\
13 & & 14.34 $\pm$ 0.04 & 11.79 $\pm$ 0.03 &11.06 $\pm$ 0.03 & 10.67 $\pm$ 0.02 & 10.25 $\pm$ 0.02 & 10.18 $\pm$ 0.02 & 10.13 $\pm$ 0.03 & 10.15 $\pm$ 0.04 & & & & \\
16 & 18.02 $\pm$ 0.05 & 15.37 $\pm$ 0.06 & 12.36 $\pm$ 0.03 &11.45 $\pm$ 0.02 & 11.03 $\pm$ 0.02 & 10.55 $\pm$ 0.02 & 10.40 $\pm$ 0.02 & 10.39 $\pm$ 0.03 & 10.39 $\pm$ 0.04 & 9.82 $\pm$ 0.18 & & & \\
17 & & 14.72 $\pm$ 0.05 & 11.95 $\pm$ 0.02 &10.74 $\pm$ 0.02 & 10.05 $\pm$ 0.02 & 9.49 $\pm$ 0.02 & 8.92 $\pm$ 0.02 & 8.46 $\pm$ 0.03 & 7.59 $\pm$ 0.04 & 3.70 $\pm$ 0.06 & 1552 & & \\
18 & & & 16.05 $\pm$ 0.11 &14.87 $\pm$ 0.07 & 14.13 $\pm$ 0.07 & 11.98 $\pm$ 0.02 & 11.37 $\pm$ 0.02 & 10.71 $\pm$ 0.03 & 9.87 $\pm$ 0.03 & 7.21 $\pm$ 0.05 & & & \\
19 & 15.10 $\pm$ 0.05 & 13.52 $\pm$ 0.05 & 9.47 $\pm$ 0.02 & 7.79 $\pm$ 0.05 & 6.46 $\pm$ 0.03 & & & 3.72 $\pm$ 0.03 & 3.04 $\pm$ 0.03 & & 71 & & \\
20 & & 17.27 $\pm$ 0.12 & 15.05 $\pm$ 0.11 &13.80 $\pm$ 0.10 & 13.14 $\pm$ 0.07 & 11.22 $\pm$ 0.02 & 10.35 $\pm$ 0.02 & 9.61 $\pm$ 0.07 & 8.70 $\pm$ 0.04 & 4.85 $\pm$ 0.04 & $<$290 & & \\
21 & & 10.95 $\pm$ 0.04 & 8.71 $\pm$ 0.03 & 7.21 $\pm$ 0.08 & 6.08 $\pm$ 0.05 & & & 3.70 $\pm$ 0.03 & 3.15 $\pm$ 0.04 & & 1501& 407.9 & \\
22 & & 14.72 $\pm$ 0.05 & 11.77 $\pm$ 0.03 &10.26 $\pm$ 0.02 & 9.44 $\pm$ 0.02 & 8.18 $\pm$ 0.02 & 7.52 $\pm$ 0.02 & 6.84 $\pm$ 0.03 & 6.04 $\pm$ 0.04 & 3.23 $\pm$ 0.04 & $<$290 & & \\
23 & & 12.39 $\pm$ 0.04 & 10.56 $\pm$ 0.02 & 9.64 $\pm$ 0.02 & 9.24 $\pm$ 0.02 & 8.70 $\pm$ 0.02 & 8.41 $\pm$ 0.02 & 8.10 $\pm$ 0.03 & 7.55 $\pm$ 0.03 & 4.58 $\pm$ 0.04 & & & \\
24 & & 13.97 $\pm$ 0.05 & 11.73 $\pm$ 0.02 &11.11 $\pm$ 0.02 & 10.78 $\pm$ 0.03 & 10.37 $\pm$ 0.02 & 10.23 $\pm$ 0.02 & 10.26 $\pm$ 0.03 & 10.26 $\pm$ 0.04 & 10.00 $\pm$ 0.09 & & & \\
25 & & 15.61 $\pm$ 0.06 & 12.04 $\pm$ 0.02 &10.72 $\pm$ 0.02 & 10.03 $\pm$ 0.02 & 9.60 $\pm$ 0.02 & 9.43 $\pm$ 0.02 & 9.42 $\pm$ 0.03 & 9.42 $\pm$ 0.04 & 9.19 $\pm$ 0.10 & & & \\
26 & 20.06 $\pm$ 0.05 & 17.27 $\pm$ 0.05 & 13.86 $\pm$ 0.03 &12.89 $\pm$ 0.03 & 12.27 $\pm$ 0.02 & 11.45 $\pm$ 0.02 & 11.02 $\pm$ 0.02 & 10.57 $\pm$ 0.03 & 9.75 $\pm$ 0.03 & 7.05 $\pm$ 0.04 & & & \\
27 & & 13.84 $\pm$ 0.05 & 11.15 $\pm$ 0.02 & 9.95 $\pm$ 0.02 & 9.17 $\pm$ 0.02 & & 8.13 $\pm$ 0.02 & & 6.37 $\pm$ 0.04 & 2.96 $\pm$ 0.04 & 44.95 & & \\
28 & 16.65 $\pm$ 0.05 & 14.50 $\pm$ 0.05 & 12.04 $\pm$ 0.02 &11.10 $\pm$ 0.02 & 10.75 $\pm$ 0.02 & & 10.26 $\pm$ 0.02 & & 10.22 $\pm$ 0.05 & & & & \\
29 & & 17.34 $\pm$ 0.12 & 15.95 $\pm$ 0.09 &15.06 $\pm$ 0.09 & 14.58 $\pm$ 0.10 & 14.21 $\pm$ 0.03 & 13.76 $\pm$ 0.03 & 13.23 $\pm$ 0.06 & 12.50 $\pm$ 0.05 & 7.12 $\pm$ 0.04 & 72 & & \\
30 & & 14.24 $\pm$ 0.04 & 11.98 $\pm$ 0.05 &11.09 $\pm$ 0.05 & 10.65 $\pm$ 0.04 & 10.06 $\pm$ 0.04 & 9.84 $\pm$ 0.04 & 9.54 $\pm$ 0.04 & 8.96 $\pm$ 0.04 & 6.42 $\pm$ 0.04 & & & \\
31 & & 15.47 $\pm$ 0.05 & 12.76 $\pm$ 0.03 &11.97 $\pm$ 0.03 & 11.49 $\pm$ 0.02 & 11.01 $\pm$ 0.04 & 10.89 $\pm$ 0.04 & 10.79 $\pm$ 0.06 & 10.57 $\pm$ 0.08 & & & & \\
32 & & 11.83 $\pm$ 0.05 & 10.48 $\pm$ 0.03 & 9.75 $\pm$ 0.02 & 9.55 $\pm$ 0.02 & 9.35 $\pm$ 0.02 & 9.28 $\pm$ 0.02 & 9.26 $\pm$ 0.03 & 9.24 $\pm$ 0.03 & 9.06 $\pm$ 0.05 & & \\
33 & & 17.25 $\pm$ 0.12 & 15.80 $\pm$ 0.07 &14.97 $\pm$ 0.08 & 14.61 $\pm$ 0.11 \\
\enddata
\tablenotetext{a}{\citet{lopezmarti04}}
\tablenotetext{b}{\citet{lopezmarti04}; Second DENIS Release.}
\tablenotetext{c}{2MASS Point Source Catalog.}
\tablenotetext{4}{\citet{haisch+04}, labeled as Cha I T14a}
\tablecomments{Data at $\lambda>3.6\mu$m are from\citet{luhman08b, luhmanmuench08, belloche11, henning93} ; this paper.}
\end{deluxetable}

In Table~\ref{Tab:params} we present a compilation of the main physical parameters of the stellar sources (spectral type, effective temperature, extinction and bolometric luminosity) reported in the literature, together with the equivalent width of the $H\alpha$ line, accretion luminosity and mass accretion rate, estimated as follows. 
We have computed for each star the photospheric continuum in the F656N filter and subtracted it from the measured flux. 
The continuum has been evaluated performing synthetic
photometry. The set of \emph{BT-Settle} synthetic spectra of
\citet{allard10} was first interpolated at the $T_{\rm eff}$ of each
source and then reddened using the $A_J$ from
\citet{luhman07}. For each spectrum we computed the photospheric $(F547M-F656N)$ color; then
using the measured F547M magnitude as a reference, we rescaled the photometry
to observed fluxes to derive the photospheric continuum 
F656N$_0$. The Equivalent Width (E.W.) of the $H_\alpha$ excess
was then derived from the ratio between the flux excess and the continuum, 
multiplied by the equivalent width of the F656N filter profile.
We converted the $H_\alpha$ excess in units of stellar luminosity by estimating
the fraction of stellar (photospheric) flux entering in the
F656N filter window, using once again the BT-Settle spectra. Then, knowing
the bolometric luminosity of the sources from \citep[from][see Table 4]{luhman07},  
we derived the $H_\alpha$ excess in units of solar luminosity
$L_\odot$. The $H_\alpha$ luminosity can be then related to the overall
accretion luminosity $L_{\rm accr}$ assuming the formula from 
\citet{demarchi10}:
\begin{equation}
\log L_{\rm accr}= 1.72 + \log L_{H\alpha}
\end{equation}
and therefore the mass accretion rates using the 
relationship \citet{gullbring98}:
\begin{equation}
L_{\rm accr} \simeq 0.8\cdot \frac{GM_{*}\dot{M_{\rm acc}}}{R_{*}}
\end{equation}
under the assumption that the mass infall onto the stellar surface starts from a distance
of about 5 stellar radii.
\begin{deluxetable}{cccccccc}
\tabletypesize{\footnotesize}
\tablecaption{Source Physical Parameters\label{Tab:params}}
\tablewidth{0pt}
\tablehead{
\colhead{N.} & 
\colhead{Sp. Type\tablenotemark{a}} & 
\colhead{T$_{eff}$\tablenotemark{a}} & 
\colhead{A$_J$\tablenotemark{a}} & 
\colhead{L$_{bol}$\tablenotemark{a}} & 
\colhead{ H$\alpha$ excess E.W.} & 
\colhead{ $\log L_{acc}$ } & 
\colhead {$\log$~\.M$_{acc}$ } 
\\
& & [K] & [mag] & [L$_\sun$] & [\AA] & [L$_\sun$] & [M$_\sun$/yr]
}
\startdata
1 & M2.75	& 3451 & 0.63 & 0.032 	& 11.9$\pm$23.5 	& -3.14$\pm$0.47 & -10.3$\pm$0.5 \\ 
2 & M3.5 		& 3342 & 0.79 & 0.42 	& 8.5$\pm$9.4 	& -2.23$\pm$0.32 &   -8.8$\pm$0.3 \\ 
3 & M0 		& 3850 & 0.79 & 0.42 	& 122.8$\pm$10.8 & -0.85$\pm$0.04 &   -7.8$\pm$0.1 \\ 
4 & M3.25	& 3379 & 0.34 & 0.33 	& 47.5$\pm$5.8 	& -1.58$\pm$0.05 &   -8.2$\pm$0.1 \\
5 & M4.5 		& 3198 & 0 	& 0.081 	& 75.0$\pm$24.6	& -2.04$\pm$0.12 &   -8.8$\pm$0.1 \\ 
6 & K5 		& 4350 & 0.45 & 0.95 	& & & \\ 
7 & K7 		& 4060 & 0.45 & 0.013 	& 124.4$\pm$10.8 & -2.23$\pm$0.04 &  -10.0$\pm$0.1 \\ 
9 & M3 		& 3415 & 1.6 	& 0.21 	& 0.9$\pm$4.3 	& -3.49$\pm$0.77 &  -10.3$\pm$0.8 \\
10 & M5.25 	& 3091 & 1.01 & 0.052	& 39.6$\pm$26.1 	& -2.56$\pm$0.22 &    -9.3$\pm$0.2 \\
11 & K6	 	& 4205 & 1.08 & 1.1 	& 19.2$\pm$11.2 	& -1.05$\pm$0.20 &    -8.1$\pm$0.2 \\
12 & M3.75 	& 3306 & 0.29 & 0.15 	& 22.3$\pm$9.7	& -2.28$\pm$0.16 &    -9.0$\pm$0.2 \\
13 & M5.5	& 3058 & 0.32 & 0.083	& 24.3$\pm$17.4 	& -2.63$\pm$0.23 &    -9.2$\pm$0.2 \\
16 & M5.75 	& 3024 & 0.5 	& 0.058	& 52.0$\pm$7.4 	& -2.44$\pm$0.06 &    -9.0$\pm$0.1 \\
17 & M3.25 	& 3379 & 1.01 & 0.15 	& 140.6$\pm$38.9 & -1.39$\pm$0.11 &    -8.3$\pm$0.1 \\
18 & M3 		& 3415 & 0.79 & 0.0029 	& 21.7$\pm$20.8 	& -3.92$\pm$0.29 &  -11.6$\pm$0.3 \\
19 & K5 		& 4350 & 1.47 & 3 		& 202.1$\pm$27.6 &   0.44$\pm$0.05 &    -6.5$\pm$0.1 \\
20 & M1.75 	& 3596 & 1.24 & 0.013	& 19.7$\pm$26.0 	& -3.29$\pm$0.36 &  -10.8$\pm$0.4 \\
21 & K5 		& 4350 & 1.35 & 5.5 	& 91.4$\pm$6.0 	&   0.42$\pm$0.03 &    -6.6$\pm$0.1 \\
22 & K8 		& 3955 & 2.14 & 0.66 	& 13.7$\pm$17.0 	& -1.30$\pm$0.35 &    -8.7$\pm$0.3 \\
23 & M1 		& 3705 & 0.54 & 0.43 	& 10.2$\pm$7.0 	& -1.91$\pm$0.23 &    -8.9$\pm$0.2 \\
24 & M5.5 	& 3058 & 0 	& 0.066 	& 28.5$\pm$14.8 	& -2.65$\pm$0.18 &    -9.2$\pm$0.2 \\
25 & M4 		& 3270 & 1.91 & 0.32 	& 19.5$\pm$19.4 	& -2.02$\pm$0.30 &    -8.6$\pm$0.3 \\
26 & M6 		& 2990 & 0.97 & 0.022 	& 599.2$\pm$369.4& -1.73$\pm$0.21 &   -8.6$\pm$0.2 \\
27 & M2		& 3560 & 1.17 & 0.42 	& 49.0$\pm$18.3 	& -1.38$\pm$0.14 &    -8.2$\pm$0.1 \\
28 & M4.25 	& 3234 & 0.81 & 0.11 	& 12.9$\pm$3.8 	& -2.68$\pm$0.11 &    -9.4$\pm$0.1 \\
29 & M2.5 	& 3488 & 0.68 & 0.003 	& 98.8$\pm$40.9 	& -3.28$\pm$0.15 &  -10.9$\pm$0.2 \\
30 & M4 		& 3270 & 0.72 & 0.11 	& 136.5$\pm$43.6 	& -1.64$\pm$0.12 &    -8.4$\pm$0.1 \\
31 & M5.75 	& 3024 & 0.56 & 0.042	& 34.8$\pm$26.0 	& -2.66$\pm$0.24 &    -9.5$\pm$0.2 \\
32 & M2.75 	& 3451 & 0 	& 0.24 \\
33 & K8 		& 3955 & 0.45 & 0.0034	& 82.9$\pm$25.5 	& -3.04$\pm$0.12 & -10.99$\pm$0.11 \\
\enddata
\tablenotetext{a}{\citet{luhman07, luhmanmuench08}}
\end{deluxetable}
\section{Source Morphology as Revealed by HST Data\label{Sec:ResolvedSources}}
In this section we illustrate our findings on the only 10  sources that appear either extended, binaries, or associated with jets and Herbig-Haro objects in the immediate vicinity. The other 23 sources appear point-like in our images. Our dataset is not homogeneous because of two main factors: a) the different sampling scales of the WFPC2 pixels (about 50~mas/pixel for the PC chip, where the main targets were located, vs. 100~mas/pixel for the WF chips where most of the other sources are found); b) the presence of CTE tail, which depends on the position and brightness of the source on the chip. To mitigate the risk of misinterpreting extended features, we have indicated the direction of the CTE deferred-charge trails with an arrow in each image. For the brightest sources, i.e. Ass Cha T2-16 (\#9), CED 112 IRS4 (\#19), and WW Cha (\#21), we have also plotted on each image the isophotal contours of the corresponding PSF, for an immediate comparison. All images shown have the standard orientation with North up and East to the left. 

\subsubsection{SX Cha (\#3)}
\begin{figure}
\epsscale{1.0}
\plotone{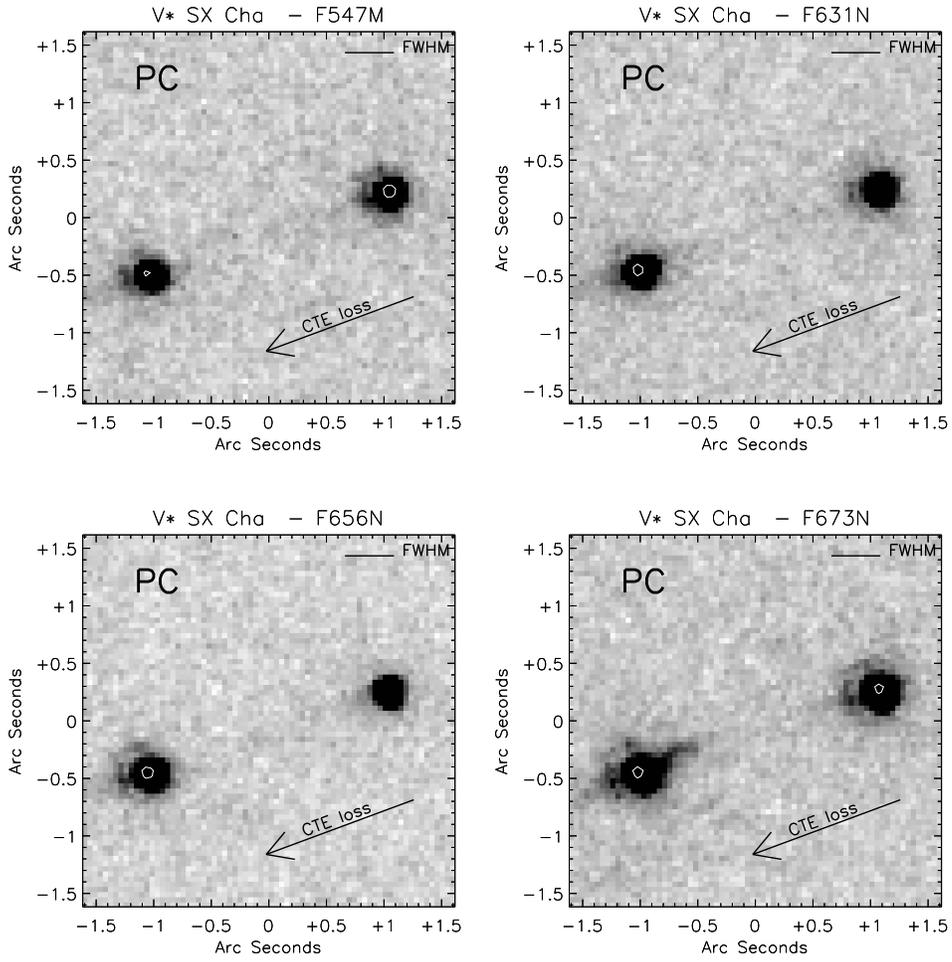} 
\caption{SX Cha observed in the PC chip (45.5 mas/pixel scale). Image centered at R.A.=10:55:59.53, Decl.=-77:24:40.3 (J2000.0).\label{T3A-1}}
\end{figure}

The T Tauri star SX Cha, spectral type M0.5 \citep{lawson96}, has a companion of comparable brightness $\sim$2$\farcs$1 at P.A.$\sim$310$\degr$ \citep{natta00}. Spitzer data show the characteristic amorphous silicate emission of circumstellar disks at 10 and 20 $\mu$m, whereas the spectral features observed in the 33-35 $\mu$m range are characteristic of crystalline enstatite and forsterite grains \citep{kessler06}.
Our HST\ images (Fig. \ref{T3A-1}) show SX Cha (source on the left) with its companion at $\sim$2$\farcs$2 (350 AU) distance. The protuberance on the western side of SX Cha visible in the F673N ([SII]) image with a length of $\sim$0$\farcs$55 (87 AU), cannot be attributed to the CTE losses. The line intensity, together with the narrow and twisted morphology, suggests that it is a collimated jet. In this case, the circumstellar disk surrounding SX Cha would be oriented somewhat perpendicularly to the plane of the binary orbit.
\subsubsection{Ass Cha T 2-5 (\#4)}
\begin{figure}
\plotone{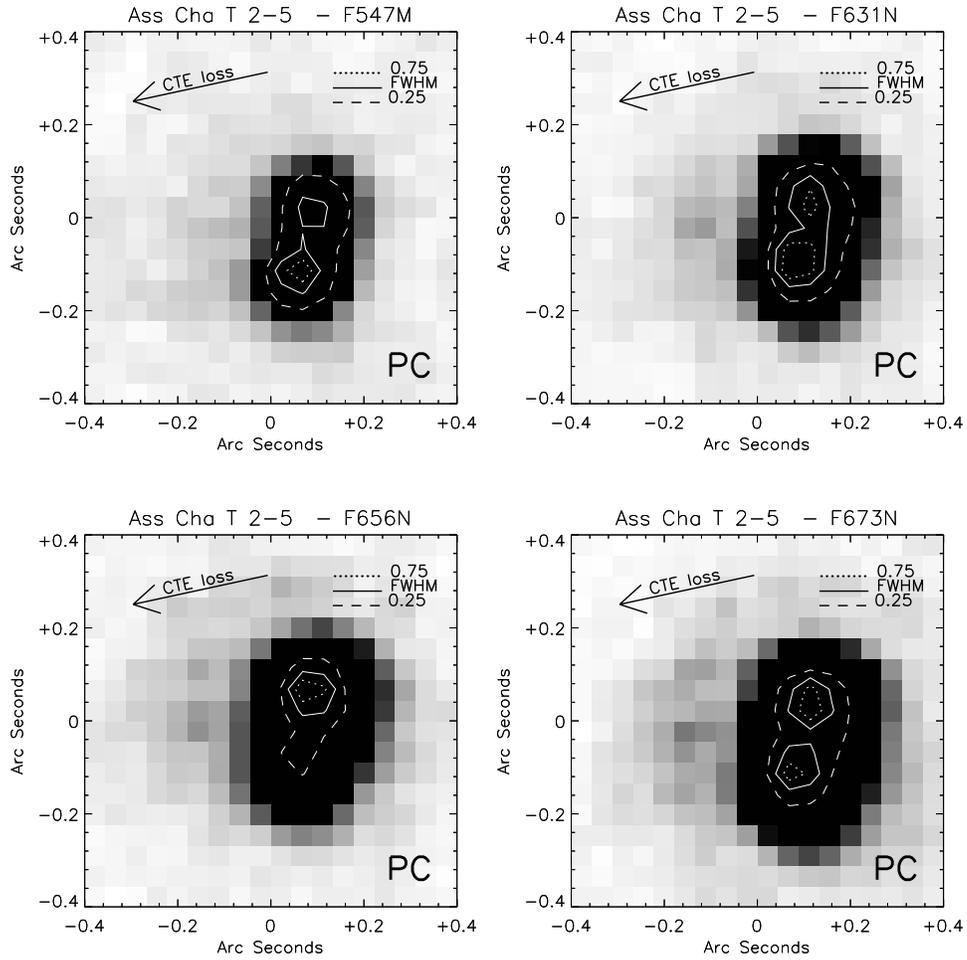} 
\caption{Ass Cha T 2-5 observed in the PC chip. The countours levels are given 0.75, 0.5 and 0.25 of the peak counts. Image centered at R.A.=10:57:42.22, Decl.=-76:59:36.4 (J2000.0). \label{T5-1}}
\end{figure}

Detected for the first time by \citet{schwartz77}, Ass Cha T 2-5 has spectral type M3.25 according to \citet{luhman07}. 
Our images (Fig. \ref{T5-1}) resolve this source in a close binary with separation $\simeq$0$\farcs$15, corresponding to $\sim$25 AU projected distance at 160pc.
The southern star appears brighter in all filters with the exception of H$\alpha$, where the northern source strongly dominates. This may indicate that the northern component has
been observed in a phase of strong accretion activity.
The ratios between the peak counts of the northern vs. the southern star are approximately 0.57 (V-band), 0.89 ([OI]), 2.8 (H$\alpha$) and 1.1 ([SII]).
\subsubsection{CT Cha (\#6)}
\begin{figure}
\plotone{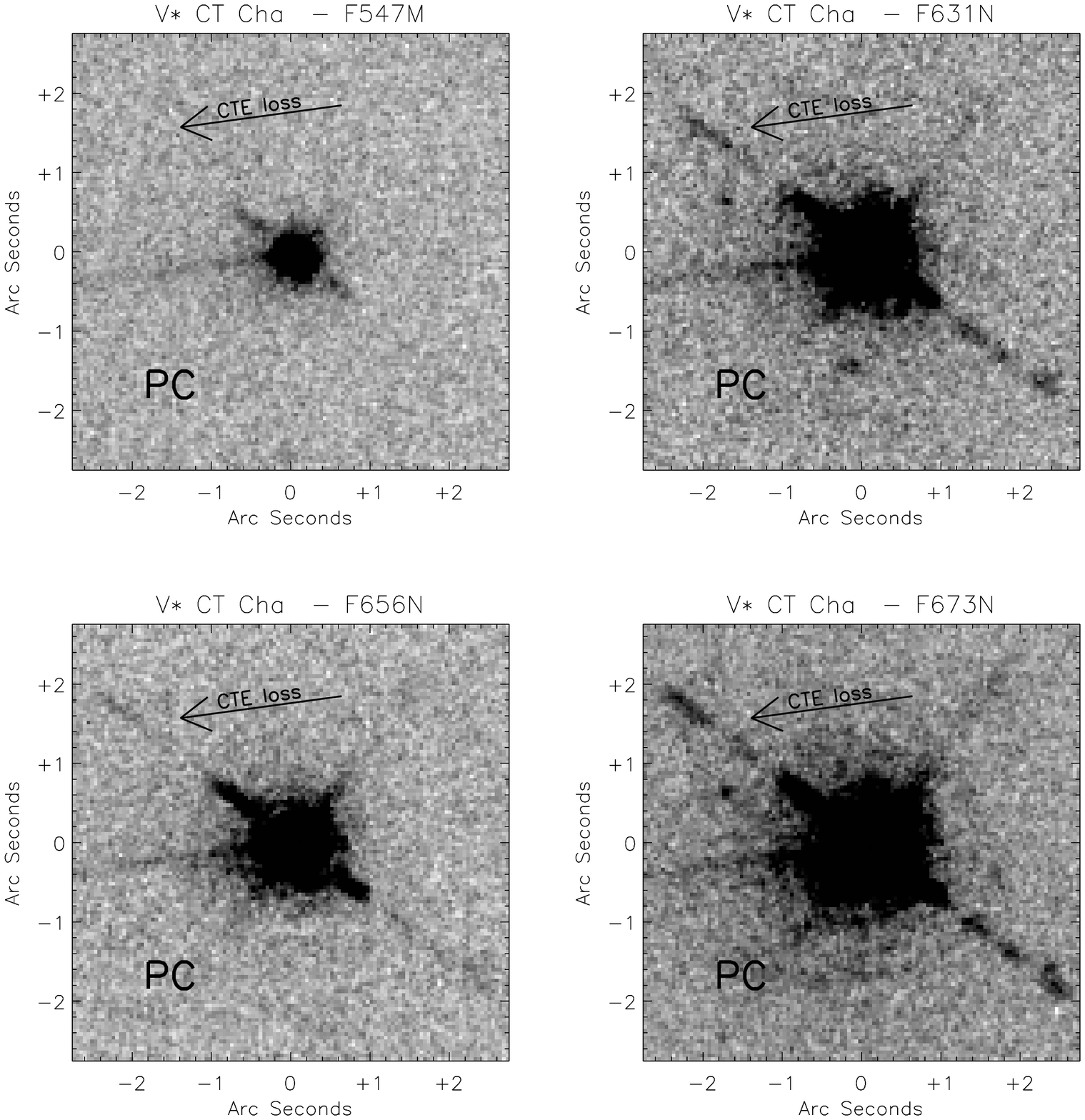} 
\caption{CT Cha observed in the PC chip. Image centered at R.A.=11:04:09.00, Decl.=-76:27:19.5 (J2000.0).\label{T14-1}}
\end{figure}

\begin{figure}
\centering
\plotone{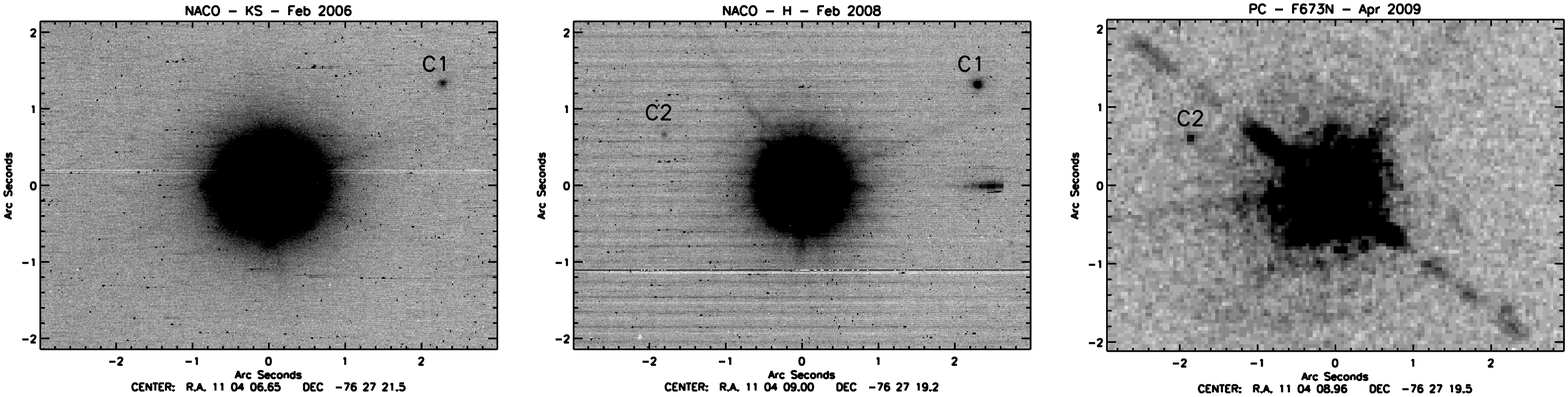}
\caption{CT Cha and the probable companions ``C1" and ``C2" (NACO's images from the ESO's archive and PC image). Our detection upper limits for source ``C1'' are: F547M: 
$m > 22.7$; 
F631N: $F_\lambda < 7.0\times10^{-19}$ erg cm$^{-2}$ s$^{-1}$ \AA$^{-1}$; 
F656N: $F_\lambda < 2.8\times10^{-18}$ erg cm$^{-2}$ s$^{-1}$ \AA$^{-1}$;
F673N: $F_\lambda < 4.6\times10^{-19}$ erg cm$^{-2}$ s$^{-1}$ \AA$^{-1}$. 
\label{T14-companion}}
\end{figure}

CT Cha has been initially classified as an emission-line star because of its H$\alpha$ line variations \citep{henizemendoza73} and later as a classical T Tauri star on the basis of its strong IRAS excess \citep{gauvinstrom92}. ISO data have shown evidence of silicate emission in a circumstellar disk \citep{natta00}. The variations of H$\alpha$ line emission have been interpreted as accretion signatures by \citet{hartmann98}, compatible with early observations of spectral veiling \citep{rydgren80}. More recent observations of CT Cha have revealed the presence of two faint companions close to the star \citep{schmidt08}. The first (source C1), about 6.3 mag fainter than the primary in the Ks-band, is located 2$\farcs$670 $\pm$ 0$\farcs$038 at P.A.$\sim$315$\degr$ of the star, corresponding to $\sim$430 AU projected distance at 160 pc. The second (C2) is about 2\arcsec\ at P.A.$\sim$45$\degr$ of the primary. \citet{schmidt08} classified C1 as physically associated with the primary due to its common proper motion, whereas for C2 they concluded that it must be a background object.
In our PC images (Fig. \ref{T14-1}) source C1 is not detected  while C2 is clearly visible at 1$\farcs$96$\pm$0$\farcs$05 projected distance (Fig. \ref{T14-companion}, right). We have found on the ESO data archive an H-band image taken with NACO in February 2006 in which only C1 is visible (Fig. \ref{T14-companion}, left). A second image taken two years later, also with NACO but in the Ks filter, shows both C1 and C2, with C2 at a distance of 1$\farcs$91$\pm$0$\farcs$03 from the primary (Fig. \ref{T14-companion}, center). Source C2 in our images is visible only in the [OI] and [SII] filters (Fig. \ref{T14-companion}, right). This, together with the fact that C2 (like source C1) has not changed position seems to indicate that it is indeed physically associated with V$^{\ast}$~CT~Cha. Its nature, however, remains enigmatic. Finally, the point-like source in the [OI] image about 1$\farcs5$ to the south of the main source may be real, as it cannot be easily attributed to a filter ghost.  
\subsubsection{HH 48 A and B (\#7, \#8)}
\begin{figure}
\plotone{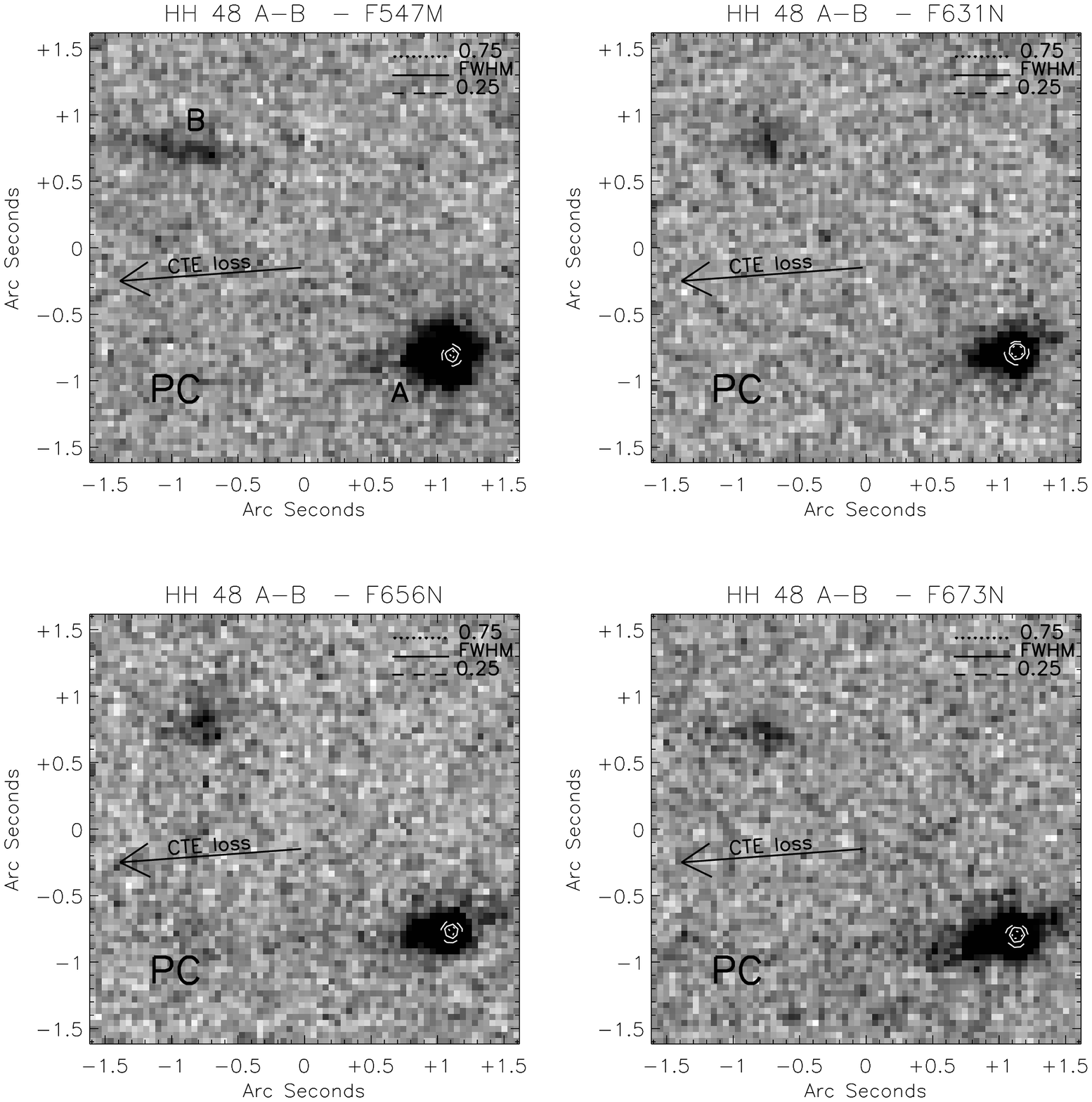} 
\caption{HH 48 A and B observed in the PC chip. Image centered at R.A.=11:04:23.04, Decl.=-77:18:08.3 (J2000.0).\label{T14A-1}}
\end{figure}

The classic Herbig-Haro object HH 48 \citep{schwartz77} is composed by two close condensations, designated as HH 48 A and HH 48 B. \citet{wanghenning06} found that HH 48 A and B are elongated in directions that are roughly perpendicular, 
while a set of newly detected features (labeled C to F) are aligned along a direction pointing to HH 48 A. This suggests the presence of two outflows driven from an embedded source, probably a binary star \citep{bally06}.
Our images (Fig. \ref{T14A-1}) show only HH\,48\,A and B, i.e. HH\,48\,C-F are not detected. In all filters, HH\,48 A appears as a bright point source associated with faint extended emission. 
While the elongation on the East side may be contaminated by CTE losses, the elongation on the West side is unambiguous. HH 48 B, located at $\sim$2$\farcs$5 (400 AU) to the northeast of HH 48 A, is fainter and elongated nearly east-west, with a length of $\simeq$0.7$\arcsec$, in the V-band filter. In the [OI] and H$\alpha$ filters, however, it appears nearly unresolved. We speculate that the morphology of HH\,48\,B is compatible with the presence of a disk seen nearly edge-on, as  the V-band elongation could be attributed to scattered light from a disk face, whereas the fainter line emission could arise from the inner disk region. Overall, our HST images confirm that HH 48 A and B point to different directions, forming an angle close to $\sim$30\degr.
\subsubsection{Ass Cha T 2-16 (\#9)}
\begin{figure}
\plotone{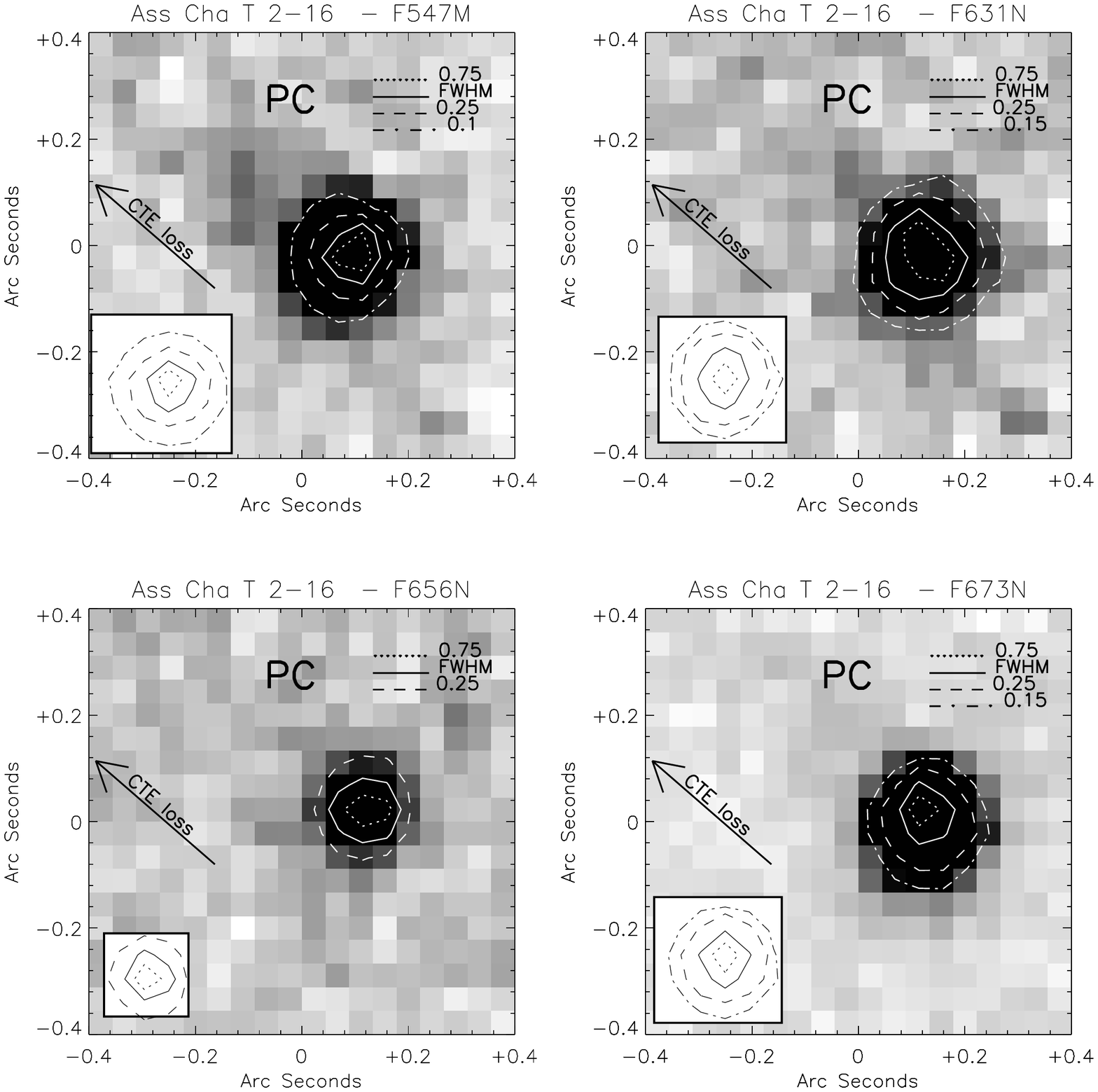} 
\caption{Ass Cha T 2-16 observed in the PC chip. Image centered at R.A.=11:04:57.01, Decl.=-77:15:57.3 (J2000.0). The insets at the bottom-left corner show the contours for the typical PSF. \label{T16-1}}
\end{figure}

This source has been classified as a M3 emission-line star with a mass of $\sim$0.26~M$_\sun$ \citep{lafreniere08}. Our images (Fig. \ref{T16-1}) indicate that Ass Cha T 2-16 has been resolved by the HST, in particular in the [OI] where the FWHM$\simeq$4 pixels is twice the size of the FWHM of unresolved sources. The SED of this source is shown in Section~\ref{sec:SEDs}.
\subsubsection{CED 112 IRS 4 (\#19)}
\begin{figure}
\plotone{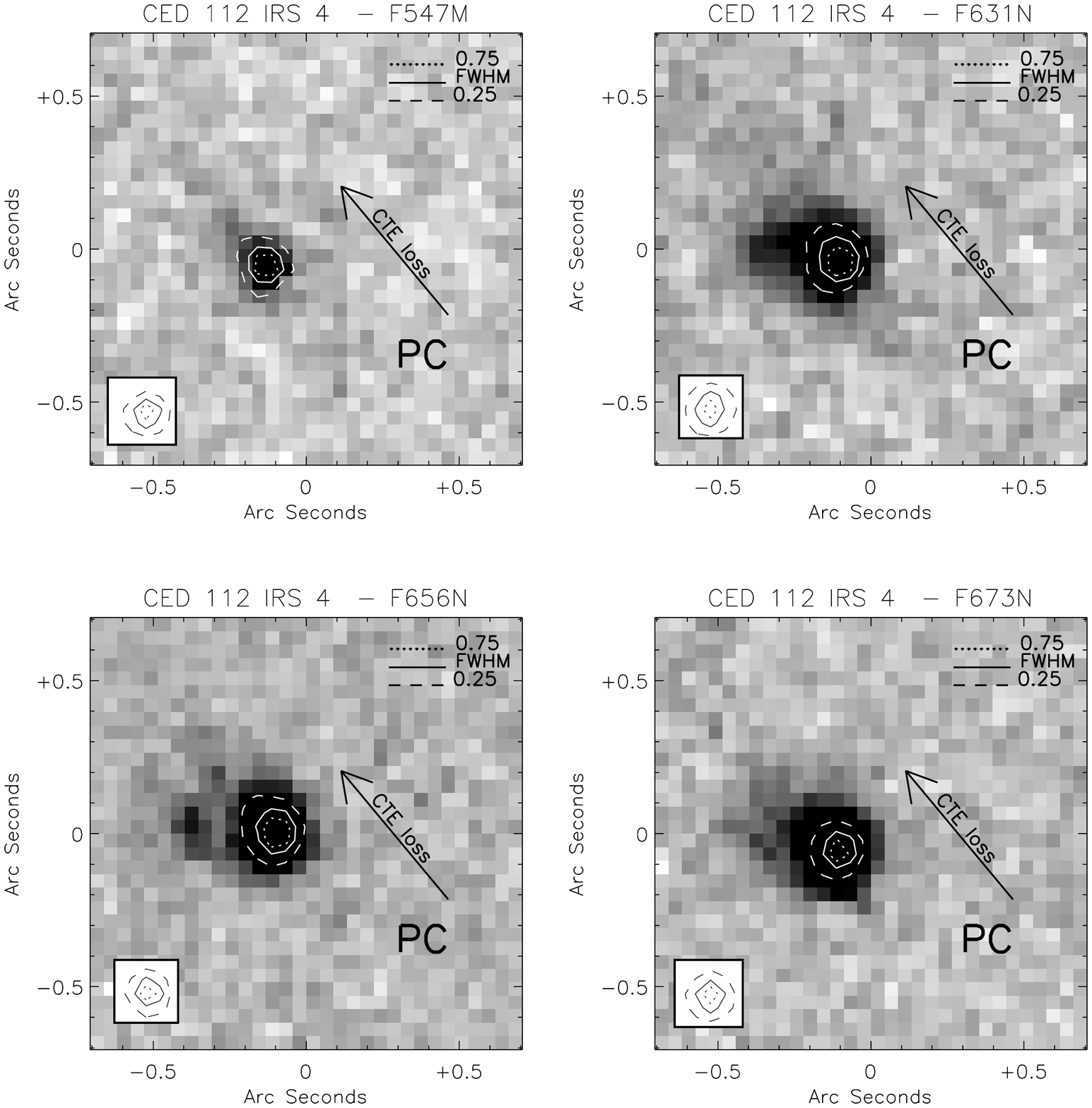} 
\caption{CED 112 IRS 4 observed in the PC chip. Image centered at R.A.=11:09:53.35, Decl.=-76:34:25.5 (J2000.0).
The insets at the bottom-left corner show the contours for the typical PSF. \label{T42-1}} 
\end{figure}

CED 112 IRS 4 is a T Tauri star located in a region rich of HH objects and circumstellar emission. \citet{wanghenning06} suggested that one of these objects,  HH 914, $\sim24\arcsec$ to the east of CED 112 IRS 4, 
is driven by it \citep[see Fig.3 of ][, where CED 112 IRS 4 is indicated as Sz 32]{wanghenning06}.
Recent ATCA\ data at 16mm show a large contribution to the 16mm flux of CED 112 IRS 4 from free-free emission \citep{lommen09}.
Our HST images (Fig. \ref{T42-1}) clearly show that the source is much brighter in the three narrow-band line filters than in the broad V-band. In particular, 
the narrow band images consistently show a feature protruding eastward, which we interpret as the HH 914 object of \citet{wanghenning06}. The H$\alpha$ image is especially remarkable, as in this filter HH 914 is most clearly detached, showing as a second peak $\sim$0$\farcs$30 (47 AU) to the East about 7.6 times fainter than the primary one.  The elongation in the V-band to the northeast is most probably due to CTE losses. The SED of this source is shown in Section~\ref{sec:SEDs}.
\subsubsection{WW Cha (\#21)}
\begin{figure}
\plotone{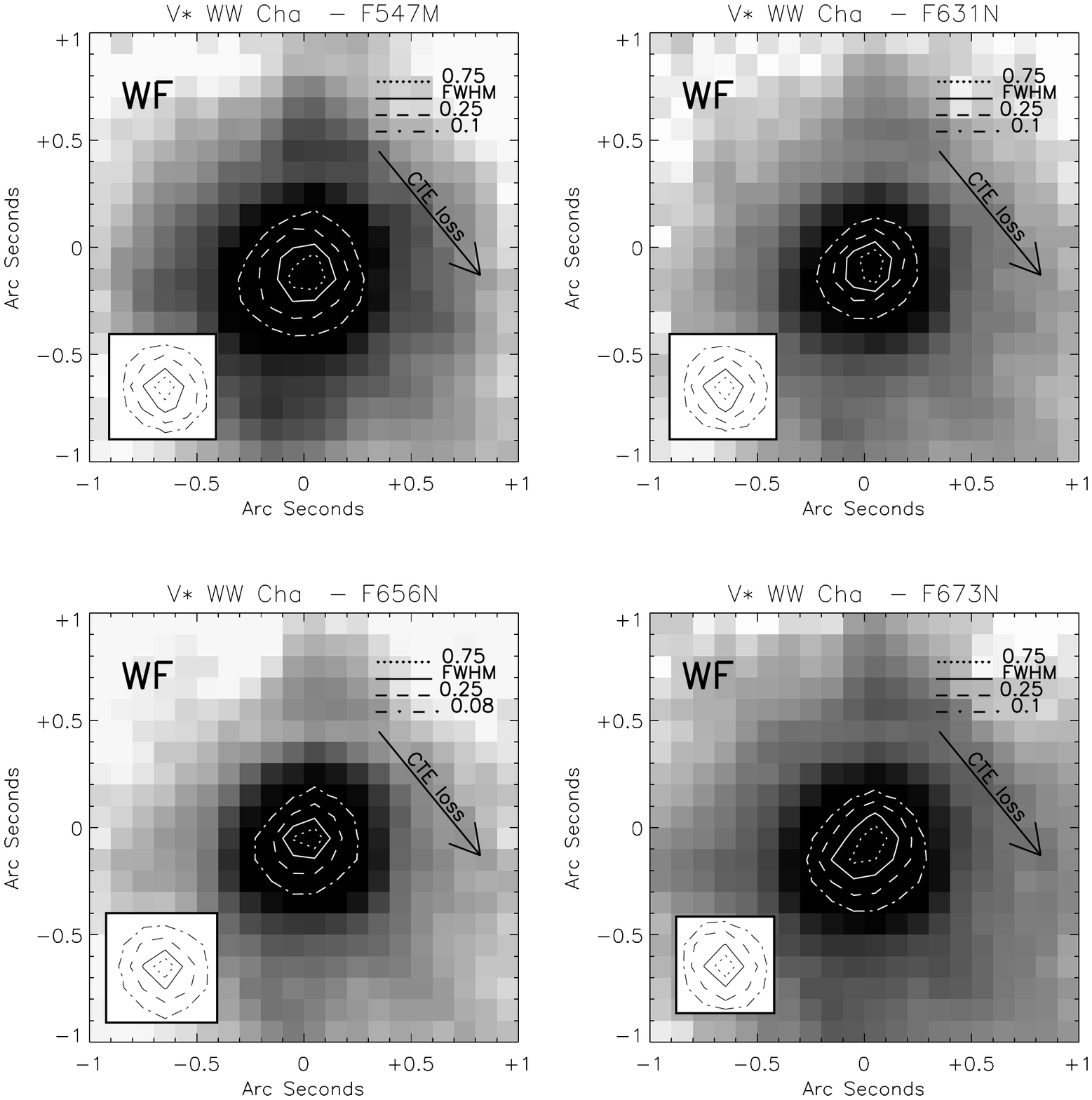} 
\caption{WW Cha observed in the WF chip. Image centered at R.A.=11:10:00.03, Decl.=-76:34:57.3 (J2000.0). The insets at the bottom-left corner show the contours for the typical PSF. \label{OTS32-2}}
\end{figure}

WW Cha, a K5 T Tauri star in the vicinity of the previous source CED 112 IRS4 \citep{schegerer06}, is thought to drive the highly collimated jets HH 915.  \citet{bally06}  report the presence of a giant bow shock, HH 931, about 13$\arcmin$ further away from WW Cha at P.A.$\sim$135$\degr$, in the same direction of HH 915. \citet{wanghenning06} suggest that two near-infrared $H_2$ emission knots, A and D, detected by \citet{gomez04} on the opposite side of WW Cha may represent the counter-jet of HH 915. There is finally a faint chain of H$\alpha$ knots which links the brightest part of the bow shock to the southern side of the reflection nebula illuminated by WW Cha. Studies of the source variability at millimeter wavelengths indicate that the 16 mm flux is dominated by cm-size pebbles' emission, which makes WW Cha the second star known to have a protoplanetary disk containing grains of suche a large size \citep{lommen09}.
Our images (Fig.~\ref{OTS32-2}) do not show evidence of the rich HH phenomenology associated to this source. Hovewer, they show the bright core extended in 
the [SII] filter, with a FWHM contour elongated in the SE-NW direction. 
If the [SII] elongation traces shock emission, then the HH 915 objects would be co-aligned and part of the same jet system.  The SED of this source is shown in Section~\ref{sec:SEDs}.
\subsubsection{ESO-H$\alpha$ 569 (\#29)}
\begin{figure}
\plotone{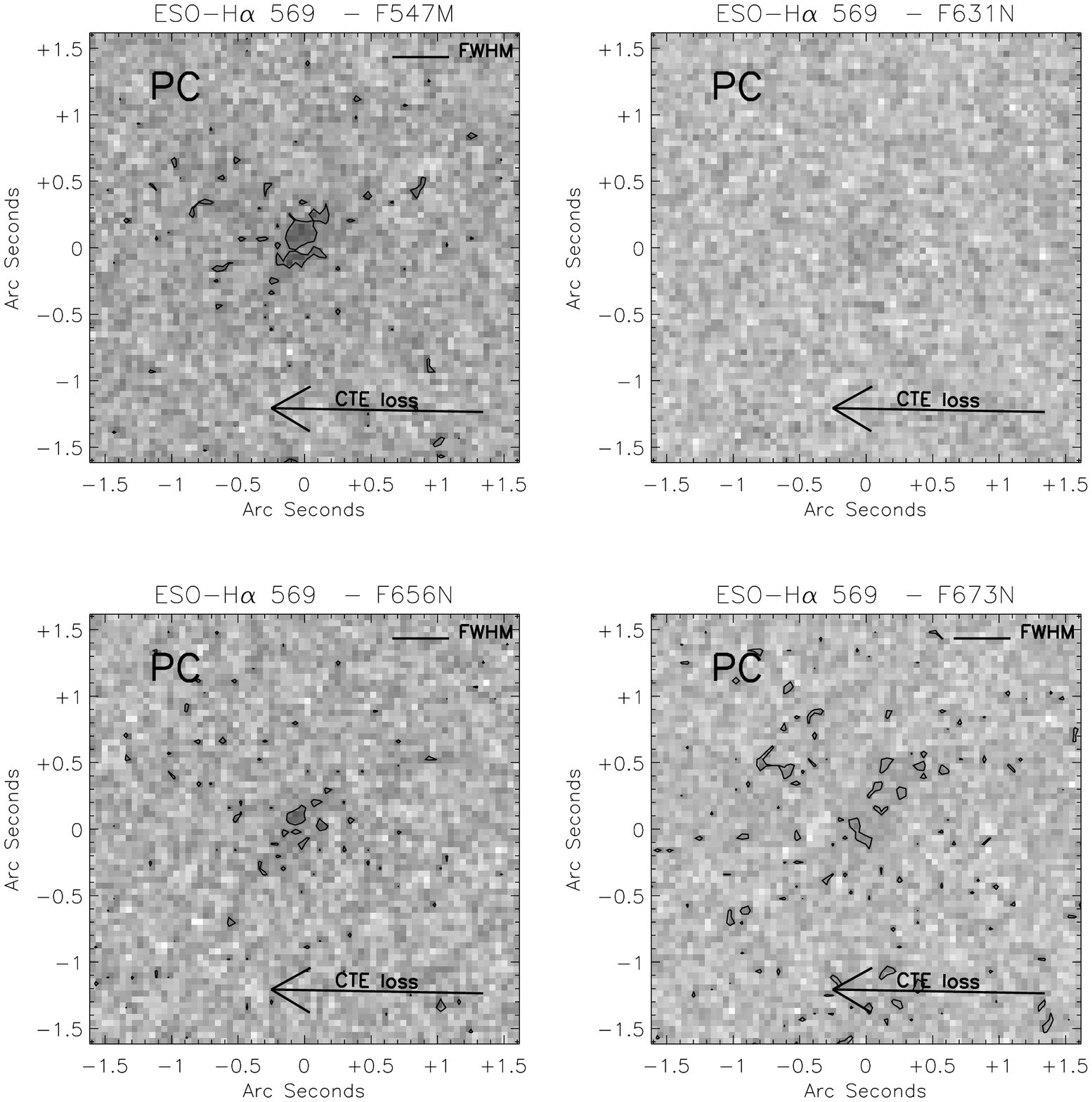} 
\caption{ESO-H$\alpha$ 569 observed in the PC chip. Image centered at R.A.=11:11:10.70, Decl.=-76:41:57.6 (J2000.0). \label{CRHF569}} 
\end{figure}

ESO-H$\alpha$ 569, spectral type M2.5, is a highly variable faint object 
with signatures of both accretion and outflow \citep{comeron04}. Also according to \citet{comeron04}, the non-detection of this source in the mid-IR by ISOCAM \citep{persi00} indicates that the amount of warm dust associated with this object is very small, meaning that its faintness cannot be attributed to occultation by circumstellar material. On the other hand, the more recent detection of this source with Spitzer \citep{luhman05} and the X-ray images of \citet{feigelson04} support the presence of an edge-on disk \citep{luhman07}. This source is located $\sim$22$\arcsec$ to the northeast of HH 919, an Herbig-Haro object that could be driven by a jet originated by ESO-H$\alpha$ 569 \citep{bally06}.
In our V-band and H$\alpha$ images, ESO-H$\alpha$ 569 (Fig. \ref{CRHF569}) is just above our detection limit, but it appears extended in the V-band in the NW-SE direction.  We do not detect HH 919, nor its associated bipolar jet.
\subsubsection{ESO-H$\alpha$ 574 (\#33)}
ESO-H$\alpha$ 574 was discovered by \citet{comeron04} as a very faint source with the characteristic colors of a lightly reddened late-type star. It has a rich emission-line spectrum dominated by the forbidden lines typically associated with stellar outflows.
\citet{comeron06} observed ESO-H$\alpha$~574 in the [SII] lines and with the low-resolution spectrograph at the ESO-VLT. They detected a well-developed jet (HH 872) protruding from the source and discussed the physical proprieties of the emitting source on the basis of the spectrum sampled at the base of the jet. Recently, \citet{bacciotti11}  have attributed the unusual low luminosity of the source in the continuum to the presence of an edge-on disk. Accounting for the disk obscuration on the luminosity of the accretion tracers, they estimate a mass accretion rate of $\sim1.7\times10^{-8}$~M$_{\sun}$~yr$^{-1}$. They also independently derive a mass outflow rate in the jet knots of about  $1.5\times10^{-9}$, leading to a mass ejection/accretion ratio over the two lobes of $\sim0.3$.  This is in the range expected for magneto-centrifugal jet launch \citep{cabrit09}.

\begin{figure}
\plotone{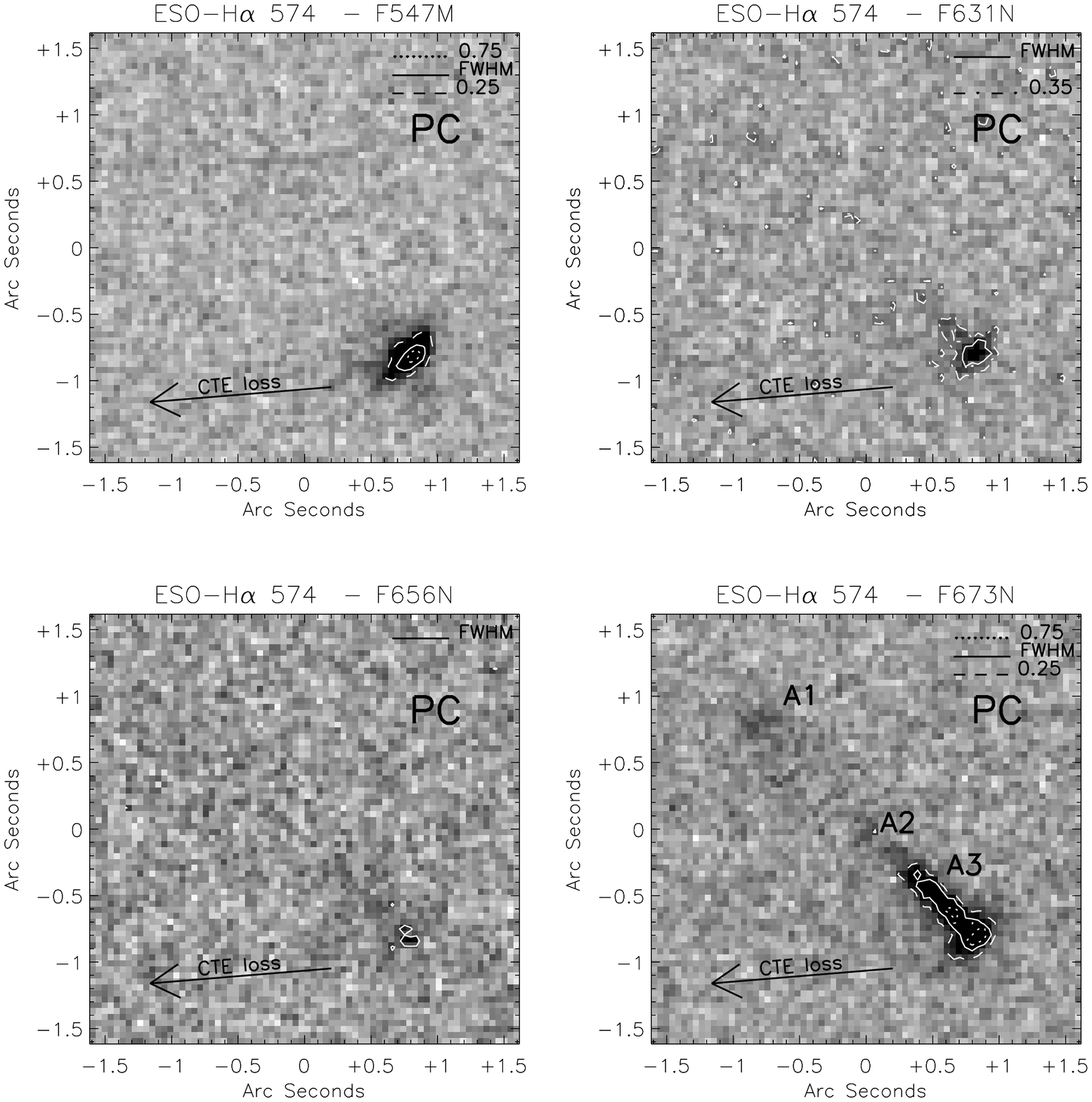} 
\caption{ESO-H$\alpha$ 574 observed in the PC chip. Image centered at R.A.=11:16:03.08, Decl.=-76:24:52.6.\label{CRHF574}}
\end{figure}

Our WFPC2 images of ESO-H$\alpha$ 574 (Fig. \ref{CRHF574}) resolve this source into a nearly edge-on disk. In the V-band the disk is bright and extends  at P.A.$\sim$135$\degr$, reaching a length of $\sim$0$\farcs$6 and a thickness, measured at the center of the disk, of $\sim$0$\farcs$4. These values correspond to 96 AU and 64 AU, respectively, assuming a distance of 160 parsec. The disk and a faint trace of a jet are also detectable in the [OI] filter, whereas in the [SII] filter the jet is clearly visible, perpendicular to the disk and extended toward the northeast direction. We resolved knot A, previously detected by \citet{comeron06}, in three knots: the brightest one, knot A3, stretches $\sim$0$\farcs$9 (144 AU) from the center of the disk.
The second knot, knot A2, is $\sim$1$\farcs$1 (176 AU) away from the disk. The third one, knot A1 \citep[previously identified by][]{bacciotti11}, is visible at a distance of $\sim$2$\farcs$3 (368 AU) from the disk and rather than being well collimated as knot A3 it appears bow-shaped. Our [SII] image does not show evidence of the southewestern counter-jet resolved by \citet{bacciotti11} in their position-velocity diagram, probably because of the relatively low sensitivity of the WFPC2/PC to diffuse structures.  The SED of this source is shown in Section~\ref{sec:SEDs}.
\section{Model fitting of the Spectral Energy Distributions \label{sec:SEDs}}
In order to estimate the main parameters of the star+disk systems 
combining the IR observations (mostly from Spitzer) with the constraints posed by our HST imaging survey we have used the online tool from \citet{robitaille06} to fit the Spectral Energy Distributions (SEDs) of a number of sources. The tool, based on a grid of 20,000 synthetic models, aims at reproducing the SEDs of disks around stars with masses between 0.1 and 50$M_{\odot}$ and ages between 1$0^3$ and 1$0^7$ years. 

Consistently with these limitations, we have fitted only the SEDs of sources having spectral class earlier than M6. We also limited ourselves to sources detected by Spitzer in at least one IRAC band, with some flux information at 24$\mu$m or longer wavelengths, typically in the sub-mm, as these data points provide a critical constrain to the IR excess from the disk. We used the photometric data reported in Tables 4 and 5, plus our own F547M photometry presented in Table 3. We excluded from the fit the sources \#4, \#7 and \#25 since these are confirmed tight binaries (see Section 4.1 and \cite{lafreniere08}). Sources \#17, \#27 and \#29 have a cluster member within the beam size of the sub-mm observations \citep{kraus07} and therefore those data points were neglected. 

The SED model fitting tool from \citet{robitaille06} aims at reproducing the photometric fluxes by varying a set of 16 parameters, of which the extinction ($A_V$) and the distance ($d$) vary in ranges defined by the user. The remaining 14 ``free'' parameters characterize each model, whose SED is computed at 10 different viewing angles. For each angle of each SED, the fitting tool calculates a $\chi^2$ to estimate the goodness of the fit. In practice, the free parameters are typically constrained by the fluxes at certain wavelengths. For example, the mass or the temperature of the star are characterized by the optical region of the SED, the inner radius or the flaring angle of the disk by the mid-far IR excess, while to constrain the disk mass one needs fluxes at $\lambda$$\gtrsim$100$\mu$m.

Given the lack of far-IR data, and the uncertainties due to the non-simultaneity of the observations, our problem is under-constrained and the SED model fitting tool returns several possible solutions. To clean-up the ensemble, for each source we considered only the solutions corresponding to values of $T_{eff}$ and $A_V$ falling within $\pm$100K and $\pm$1mag from the values given in Table 4. This is appropriate, considering that the spectral types and absolute temperature scale of low-mass stars have typical uncertainty of one sub-class. We also constrain the distance to be in the range between 130pc and 190pc, an artifice to allow for some variability of the source luminosity \citep[see e.g.][]{Morales-Calderon+11}. As the average of the calculated distances remains close to 160~pc, this does not introduce systematic trends in the derived luminosities. We finally require that the  difference between the $\chi^2$ and the best $\chi^2$ of each source must be smaller than 3. This latter criterion, also adopted by \citet{robitaille07} to fit low-mass young stellar object (YSO) SEDs, has been necessary because the model grid is too sparse to reliably search for the true minima of the $\chi^2$ hypersurface. 

\begin{figure}
\centering
\includegraphics[scale=0.85]{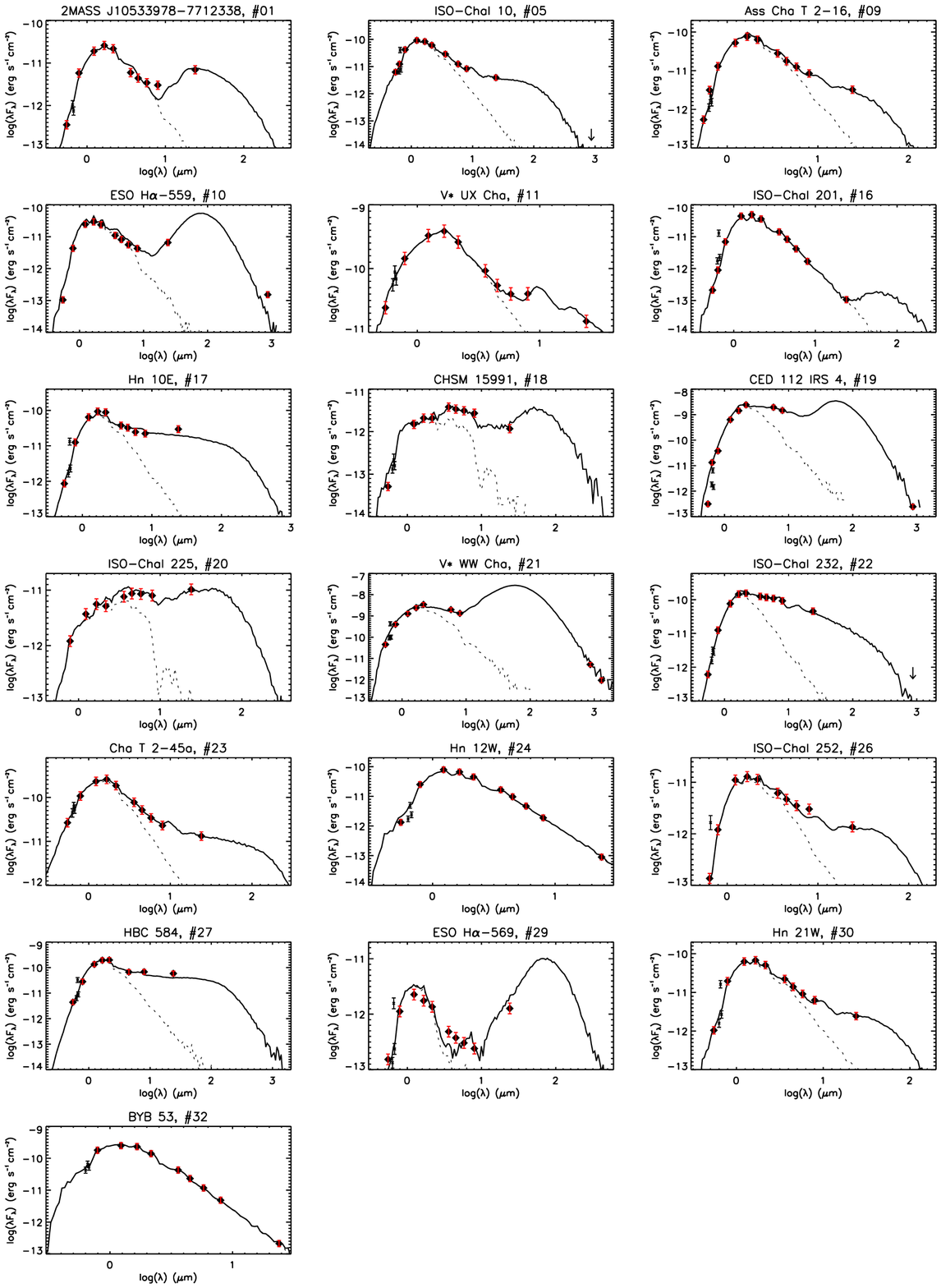}
\caption{SEDs of our sample of sources. Data obtained in our HST line-filters  have not been used for the fit as they may be contaminated by accretion or mass loss. They are presented in the plot with black error bars. The data actually used for the fit have red error bars.\label{seds}}
\end{figure}

Since the minima of the $\chi^2$ hypersurface are difficult to resolve, we could not reliably 
identify the best fit model that formally represents the physics of the source, with its associated confidence interval. 
However, as shown by \citet{grave09}, the distributions of fundamental parameters such as the stellar mass, age and total luminosity tend to show peaks independently on the finesse of the model grid. Following these authors, we have used these distribution (constrained by our knowledge of the stellar parameters) to derive the ``best fit'' parameters with their uncertainties, computing a weighted mean and a weighted standard deviation for each parameter and each source. For the weights we used the inverse of the $\chi^2$ returned by the SED model fitting tool. The mean and the standard deviation were computed on a logarithmic scale because the parameters in the grid are usually uniformly sampled in logarithmic scale. The resulting distributions of the stellar age and mass show
that when a large number of models can be found, they tend to produce a rather narrow peak in the distribution of stellar parameters. Only for three sources, \#18, \#20 and \#29, the number of models passing our selection criteria is small and  the corresponding histograms do not show a peak. Following this approach we were able to fit 19 SEDs, shown in Fig.~\ref{seds}. The physical parameters derived from the SED fitting, together with their uncertainties, are listed in Tables \ref{s_par} and \ref{d_par}.

\subsection{Parameters derived from SED fitting}
Table \ref{s_par} provides the parameters related to the central source: the entry number we assigned to each star in Table 2 (column 1); the $\chi^2$ per data-point of the best fit (column 2); the number of fits that satisfy our selection criteria on $T_{eff}$, $A_V$ and $\chi^2$ (column 3); and finally, in the last four columns, the bolometric luminosity, age, mass and radius of the central source. 
Most of the sources have sub-solar mass down to 0.12~$M_{\odot}$, a value close to the low-mass limit of the fitting tool. The median age, 2.5~Myr, is in agreement with previous estimates, while the spread between is 0.5~Myr and 5~Myr, with a few outliers.  In general, this spread is in line with what typically found using isochronal timing of PMS clusters \citep{hillenbrand+2008}. It is known that several factors may affect the estimate of the absolute stellar luminosity ofPMS stars \citep[see e.g.][]{hartmann01, reggiani+11}, and in particular edge-on disks can explain the most extreme cases of sub-luminous sources. If the age spread is real, star formation in Chamaeleon I proceeded rather slowly and may still be  ongoing, as suggested by \citet{luhman07} and \citet{belloche11}. 

Table~\ref{d_par} refers to the main disk parameters. For each source we list the entry number (column 1); the infrared spectral index $\alpha_{2-24}$, defined as $\alpha=\textit{d}\log(\lambda F_{\lambda})/\textit{d}\log\lambda$ \citep{lada84}, calculated between 2.2$\mu$m and 24$\mu$m and dereddened using the extinction from Table 4 and the reddening law from \citet{flaherty07} (column 2); the mass of the disks, either derived by the fitting tool 
(column 3) or directly estimated from the data at 870$\mu$m (column 4); the dust sublimation radius $R_{sub}=R_*(T_*/T_{sub})^{2.1}$ \citep{robitaille06} for a dust sublimation temperature, $T_{sub}=1600$~K (column 65;  the inclination of the disk to the line of sight (column 6).  The SED fitting tool also estimates disk scale height factors $z_{factor}\sim0.8$ and disk flaring parameters $\beta\sim1.0$. These values turn out to be similar for all sources within about 10$\%$.

In Table~\ref{d_par} we provide the model disk mass $M_d^{SED}$ only for sources with some flux measure at $\lambda$$\gtrsim$100$\mu$m, which allows constraining the amount of dust in the outer disk regions where most of the mass resides. These estimates can be compared with the the disk mass estimated from the flux density of sub-mm continuum dust emission (column 4), calculated assuming optically thin emission at long-wavelengths. The flux density in the sub-mm, $F_{\rm{sub-mm}}$, can be converted into an estimate of the disk mass through the relation
$M_{\rm{disk}} \simeq d^2 F_{\rm{sub-mm}} /(\kappa_{\nu} B_{\nu} (T_c))$, where $d$ is the distance (160~pc), $\kappa_{\nu}$ is the total (gas + dust) mass opacity, $B_{\nu} (T_c)$ is the Planck function at the characteristic temperature $T_c$ of the emitting dust. For the mass opacity we adopted the opacity law of \citet{beckwith90}, i.e. $\kappa_{\nu}=0.1 \times (\nu/10^{12}$Hz$)$~cm$^2/$g, whereas we considered a value of 20~K for the characteristic outer disk temperature \citep[see e.g.][]{andrews05}. 
Unfortunately, the sub-mm data can be reliably used to derive the disk mass only for 3 sources, as in the other cases we have either a non-detection or a cluster member, according to the lists of \citet{kraus07} and \citet{lafreniere08}, close enough to potentially affect the measured sub-mm flux. The values obtained in those cases can be regarded as upper limits to the real disk mass, and are therefore reported in square brackets. 


\begin{deluxetable}{cccccccccc}
\rotate
\tablecaption{Stellar parameters obtained from the fit\label{s_par}}
\tablewidth{0pt}
\tablehead{
\colhead{N.} &
\colhead{$\chi^2$ best fit} & 
\colhead{Accepted Fits} & 
\colhead{$L_{bol}$} &
\colhead{Age} &
\colhead{$M_*$} &
\colhead{$R_*$} \\
& & & [$L_{\odot}$] & [log yr] & [$M_{\odot}$] & [$R_{\odot}$]
}
\startdata
1 & 1.1 & 242 & 0.09 $\pm$ 0.03 & 6.81 $\pm$ 0.13 & 0.28 $\pm$ 0.03 & 0.81 $\pm$ 0.12 \\
5 & 0.087 & 818 & 0.10 $\pm$ 0.05 & 6.57 $\pm$ 0.13 & 0.19 $\pm$ 0.03 & 1.0 $\pm$ 0.2 \\
9 & 0.62 & 768 & 0.18 $\pm$ 0.06 & 6.49 $\pm$ 0.14 & 0.31$\pm$ 0.03 & 1.2 $\pm$ 0.2 \\
10 & 4.3 & 12 & 0.2 $\pm$ 0.3 & 5.9 $\pm$ 0.6 & 0.17 $\pm$ 0.02 & 1.5 $\pm$ 0.8 \\
11 & 0.055 & 1524 & 0.8 $\pm$ 0.3 & 6.5 $\pm$ 0.2 & 0.93 $\pm$ 0.07 & 1.7 $\pm$ 0.3 \\
16 & 0.46 & 1584 & 0.08 $\pm$ 0.03 & 6.4 $\pm$ 0.3 & 0.12 $\pm$ 0.02 & 1.0 $\pm$ 0.2 \\
17 & 0.80 & 226 & 0.28 $\pm$ 0.13 & 6.4 $\pm$ 0.2 & 0.30 $\pm$ 0.03 & 1.4 $\pm$ 0.4 \\
18 & 0.33 & 6 & 0.6 $\pm$ 0.4 & 6.0 $\pm$ 0.3 & 0.31 $\pm$ 0.02 & 2.1 $\pm$ 0.7 \\
19 & 3.4 & 33 & 25 $\pm$ 14 & 5.2 $\pm$ 0.2 & 1.7 $\pm$ 0.4 & 9 $\pm$ 2 \\
20 & 0.74 & 6 & 1.2 $\pm$ 0.3 & 5.8 $\pm$ 0.2 & 0.40 $\pm$ 0.03 & 2.8 $\pm$ 0.5 \\
21 & 0.44 & 90 & 20 $\pm$ 12 & 5.3 $\pm$ 0.3 & 1.6 $\pm$ 0.4 & 7 $\pm$ 3 \\
22 & 0.27 & 54 & 1.2 $\pm$ 0.4 & 6.4 $\pm$ 0.3 & 0.69 $\pm$ 0.08 & 1.6 $\pm$ 0.4 \\
23 & 0.042 & 1070 & 0.39 $\pm$ 0.16 & 6.4 $\pm$ 0.2 & 0.47 $\pm$ 0.03 & 1.4 $\pm$ 0.3 \\
24 & 0.41 & 1640 & 0.08 $\pm$ 0.04 & 6.5 $\pm$ 0.2 & 0.13 $\pm$ 0.02 & 1.0 $\pm$ 0.2 \\
26 & 0.33 & 679 & 0.06 $\pm$ 0.06 & 6.5 $\pm$ 0.4 & 0.12 $\pm$ 0.02 & 0.8 $\pm$ 0.4 \\
27 & 0.73 & 126 & 0.5 $\pm$ 0.3 & 6.2 $\pm$ 0.2 & 0.39 $\pm$ 0.03 & 1.8 $\pm$ 0.6 \\
29 & 2.0 & 5 & 0.8 $\pm$ 0.4 & 5.8 $\pm$ 0.2 & 0.33 $\pm$ 0.03 & 2.5 $\pm$ 0.6 \\
30 & 0.076 & 1602 & 0.10 $\pm$ 0.05 & 6.59 $\pm$ 0.15 & 0.23 $\pm$ 0.03 & 1.0 $\pm$ 0.2 \\
32 & 0.030 & 614 & 0.22 $\pm$ 0.09 & 6.38 $\pm$ 0.13 & 0.32 $\pm$ 0.02 & 1.3 $\pm$ 0.2 \\
\enddata
\end{deluxetable}

\begin{deluxetable}{cccccc}
\tablecaption{Disk parameters obtained from the fit\label{d_par}}
\tablewidth{0pt}
\tablehead{
\colhead{N.} &
\colhead{$\alpha_{2-24}$} & 
\colhead{$M_{d}^{SED}$} & 
\colhead{$M_{d}^{sub-mm}$\tablenotemark{a}} & 
\colhead{$R_{sub}$} & 
\colhead{Inclination} \\
& & [$M_{\odot}$] &[$M_{\odot}$] & [AU] & [deg]
}
\startdata
1 & -0.52 $\pm$ 0.12 & --- 	& --- 		& 0.02 & 81.4 \\
5 & -1.14 $\pm$ 0.12 & --- 	& $<$0.13 	&  0.02 & 75.5 \\
9 & -1.35 $\pm$ 0.12 & --- 	& --- 		& 0.03 & 75.5 \\
10 & -0.60 $\pm$ 0.12 & 	& 0.003 		& 0.03 & 87.1 \\
11 & -1.27 $\pm$ 0.27 & --- & ---  		& 0.06 & 63.3 \\
16 & -2.45 $\pm$ 0.25 & --- & --- 	 	& 0.016 & 87.1 \\
17 & -0.53 $\pm$ 0.14 & --- & [0.12] 		& 0.03 & 18.2 \\
18 & -0.29 $\pm$ 0.17 & --- & --- 		& 0.05 & 87.1 \\
19 & --- & 0.001 $\pm$ 0.004 & 0.005 	& 0.3 & 75.5 \\
20 & 0.20 $\pm$ 0.17 & --- & $<0.03$ 	& 0.8 & 87.1 \\
21 & --- 	& 0.02 $\pm$ 0.08 & 0.12 		& 0.3 & 31.8 \\
22 & -0.66 $\pm$ 0.12 & --- & [$<0.03$] 	& 0.05 & 41.4 \\
23 & -1.14 $\pm$ 0.12 & --- & --- 		& 0.04 & 69.5 \\
24 & -2.58 $\pm$ 0.17 & --- & --- 	 	& 0.016 & 81.4 \\
26 & -0.95 $\pm$ 0.12 & --- & --- 		& 0.016 & 75.5\\
27 & -0.59 $\pm$ 0.12 & --- & [0.003] 	& 0.05 & 18.2 \\
29 & -0.08 $\pm$ 0.19 & --- & [0.005] 	& 0.05 & 87.1 \\
30 & -1.32 $\pm$ 0.14 & --- & --- 		& 0.02 & 75.5 \\
32 & -2.69 $\pm$ 0.13 & --- & --- 		& 0.03 & 56.6 \\
\enddata
\tablenotetext{a}{In this column we report the disk mass derived from the data at 870$\mu$m (see Section 2.3). The values relative to sources with a visual companion within the beam are given in brackets and can be considered as upper limits to the real disk masses.}
\end{deluxetable}

\subsection{Analysis of the Best-fit Model Parameters}
In this final section we discuss a number of diagrams useful to address the presence of evolutionary trends between the main parameters of our star+disk systems. 

\begin{figure}[h!]
\centering
\includegraphics[scale=0.5]{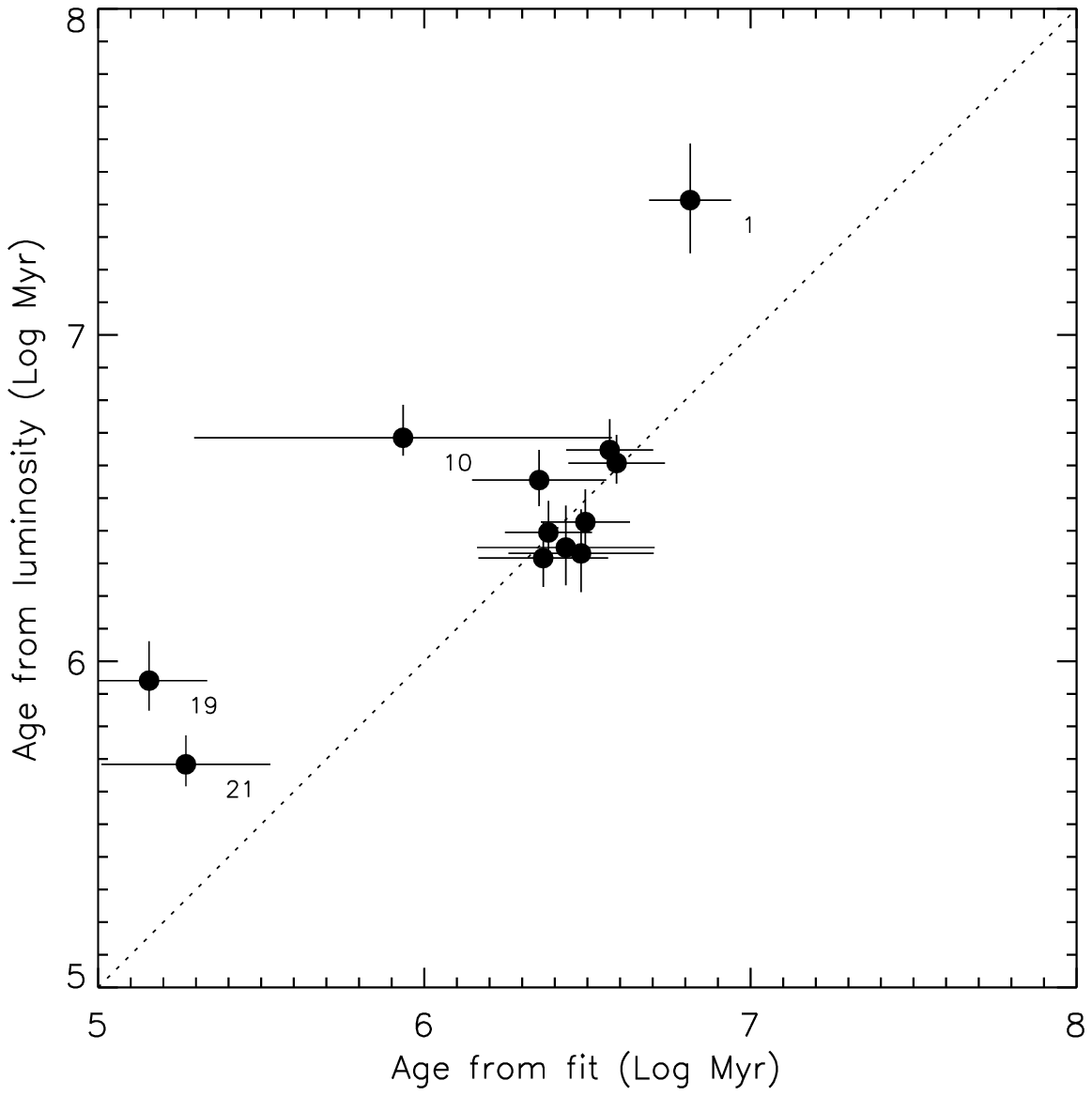}
\caption{Comparison between the ages estimated with Luhman luminosities + Siess tracks and the SED model fitting tool from \citet{robitaille06}.\label{age_RobSiess}}
\end{figure}

\begin{figure}[h!]
\centering
\includegraphics[scale=0.5]{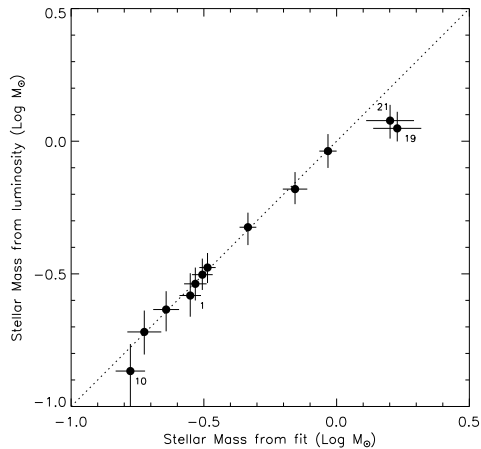}
\caption{Comparison between the stellar masses estimated with Luhman luminosities + Siess tracks and the SED model fitting tool from \citet{robitaille06}.\label{mass_RobSiess}}
\end{figure}

First, as a sanity check, we compare the values of stellar age and mass returned by the fitting tool with those obtained from the direct interpolation of $T_{eff}$ and $L_{bol}$ in the HR diagram using the Siess tracks \citep{siess00}, which are the same tracks adopted by the fitting tool. Fig. \ref{age_RobSiess} and \ref{mass_RobSiess} show that there is strong agreement. 
In particular, the strong correlation between the mass values is expected, since the model fitting assumed temperatures  within $\pm100$~K from the stellar temperatures and in this temperature range the mass is strongly correlated to the temperature. There are, however,  four sources (\#1, \#10, \#19, \#21) for which the fitting tool provides ages and masses which are discrepant from those found in the literature.  The  younger ages we derive are a result of the higher bolometric luminosities returned by the fitter.  For disk seen at high inclinations, taking into account the IR part of the SED allows to recover a non-negligible fraction of the stellar flux which would otherwise remain unaccounted for by a applying a simple reddening correction to the optical photometry.   \citet{luhman04} and \citet{luhman07}, who derived $L_{bol}$ from I and J band magnitudes for almost the entire known population of Chamaeleon~I, already noticed that sources with unusually low $L_{bol}$ values, which apparently lie below the Main Sequence in the HR diagram, may be highly affected by an underestimate of the dust column density.  Unfortunately these systems are also the most challenging to model, due to the lack of direct information e.g. on the disk flaring angle and on the dust properties at the disk surface.   
With the exception of source  \#10,  whose age (i.e. luminosity) derived from the fit is highly uncertain, 
the other three  sources 
are distributed at the two extremes of the age range: source \#1 is the most luminous while source \#19 and \#21 are the faintest ones. Our HST images show that these last two objects are associated with diffuse emission, consistent with sources seen mainly in scattered light.

\begin{figure}[h!]
\centering
\includegraphics[scale=0.5]{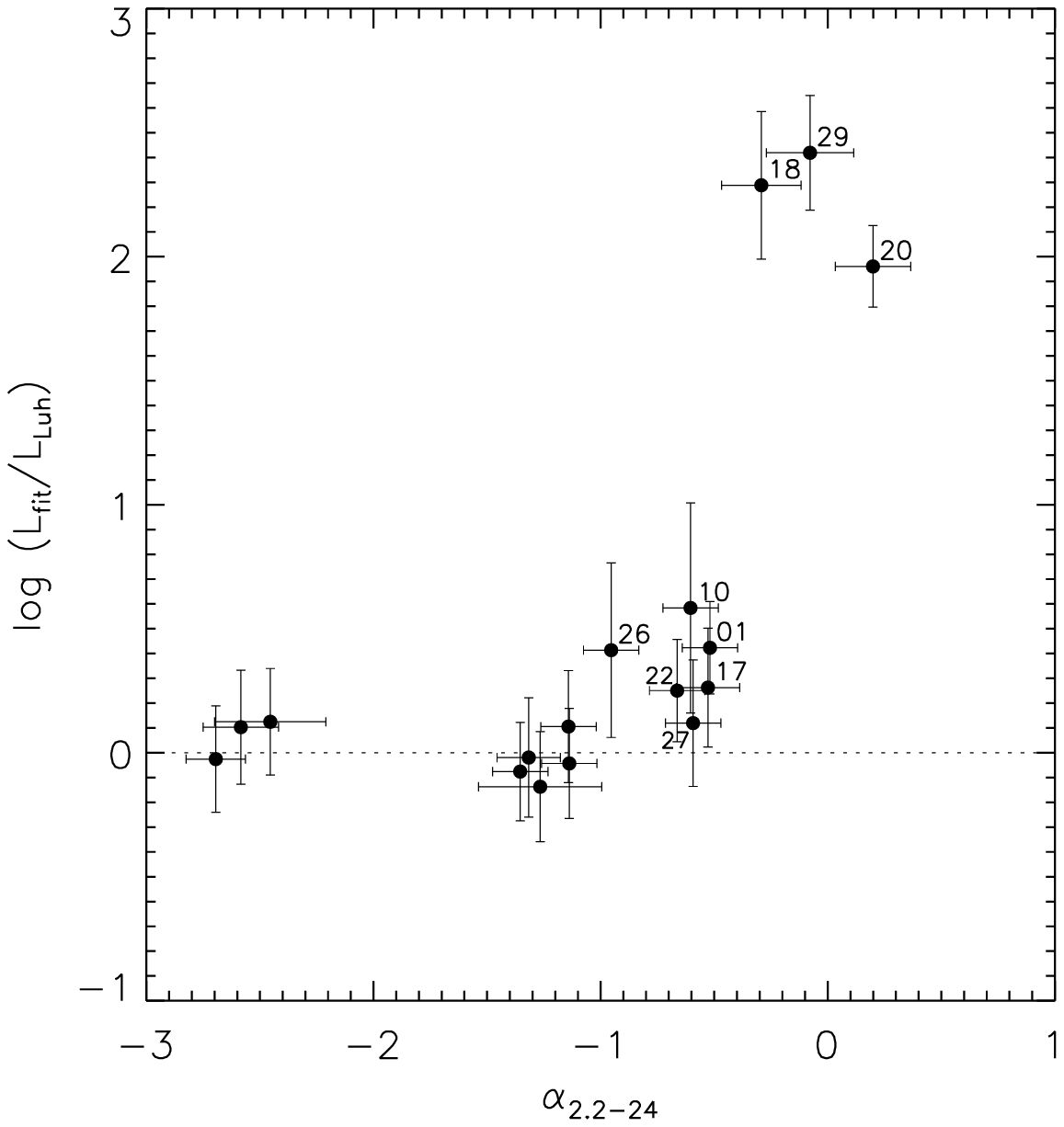}
\caption{Difference between the values of $L_{bol}$ derived in this work and those from \citet{luhman07} versus the infrared spectral index $\alpha_{2-24}$ that is a tracer of the structure of the inner disk, being mostly dependent on the flaring angle and on the presence of an inner rim.\label{nK24_difLuh}}
\end{figure}

Table~\ref{d_par} shows that 14 out of 19 disks have tilt angle within 30 degrees from edge-on, about twice the number one would expect if  disks are oriented randomly. While this may be partially due to the sparse sampling of this parameter provided by  the theoretical model,  we must remark that the model may actually be biased toward larger angles, as it may not sample disks that are enough settled. The model adopts a fully flared disk (in vertical hydrostatic equilibrium) and mimics the dust settling by multiplying the scale height at the dust sublimation radius by a ``zfactor''. Figure 6 in \citet{robitaille06} shows that the zfactor is a function of the disk outer radius, becoming equal or less than 0.5 only for relatively large disks. However, \citet{szucs+10} notice that the average SED of T-Tauri stars in the Chamaeleon-I region already requires a reduction of a factor of 2 in the disk scale height; similar results are also reported by the Spitzer Infrared Spectrograph GTO team for Taurus sources \citep{furlan+11}. This suggests that the grid of Robitaille et al. may not properly sample the typical settling of TTau disks. In conclusion, the returned values for the tilt angle are only indicative, and therefore  we use them to flag outliers without drawing any further conclusion. 

To further investigate the possible correlations between the disk structure and the derived stellar properties, we  compare the infrared spectral index $\alpha_{2-24}$, a tracer of the amount of dust in the warmest layers of the circumstellar disk, to the main parameters of the central sources, like luminosity and  mass. 
Fig.~\ref{nK24_difLuh} shows the difference between the values of $L_{bol}$ derived by the fitting tool and by \citet{luhman07} against the spectral index $\alpha_{2-24}$. Three sources with discrepant luminosity stand out in the plot: \#18, \#20 and \#29. These three sources, with nearly flat IR SEDs, were positioned according to \citet{luhman07}  below the Zero-Age-Main-Sequence and therefore, having no estimate for their mass and age, did not appear in Figures~\ref{age_RobSiess} and \ref{mass_RobSiess}. The fitting tool returns for them a small number of successes, confirming that these systems have extreme properties. In particular, these are the sources with the highest tilt angle, seen nearly edge-on. Our derived ages and mass are now in-line with the main population (see Figures~\ref{nK24_time} and \ref{nK24_massc} below).  Sources \#19 and \#21 do not appear in this plot as they have no detection at 24~$\mu$m.

Fig.~\ref{nK24_difLuh}  also shows that for the other sources the difference between the estimated luminosities appears to increase with the spectral index: there is a correlation between the amount of warm dust seen in the outer disk layers and the apparently low bolometric luminosity derived from dereddening the I and J-band magnitudes. In general, this plot seems to confirm an underestimate of the dust column density toward the edge-on sources.
 
Fig.~\ref{nK24_time} shows that the index $\alpha_{2-24}$ decreases with time for nearly all sources (\#19, and \#21, younger than 0.2~Myr and without a 24~$\mu$m measure, are not plotted), indicating that the disks become flatter with age. This is most probably related to the gradual settling of the dust grains on the disk plane. Our linear fit to the  main distribution is provided by the formula shown in the figure. \citet{luhman08b} set  $\alpha_{2-24}=-2.2$ as the lower limit for disk dissipation. According to our expression, this value is reached at a time $\simeq10^7$\,yr, compatible with the scenario that disks around {\sl most} young solar analog stars clean out their small dust grains within 1~AU in $\approx 10$~Myr or less \citep{pascuccitachibana2010}.
The three sources  that strongly deviate from the main trend, \#16, \#24 and \#32, are those exhibiting negligible IR excess. These outliers seem to have dissipated their inner disk relatively quickly, less than $\sim3$~Myr. Various effects may lead to rapid disk dissipation, all intimately linked to the physical or chemical properties of the environment, like e.g. higher UV or X ray flux from the central star, tidal forces due to the presence of a close companion or of a giant planet \citep{cieza09}, or low disk metallicity which can increment the dust photoevaporation rate \citep{ercolano10}. A larger statistical sample is needed to quantify the frequency of premature versus delayed disk dissipation.

A plot of the $\alpha_{2-24}$ index against stellar mass (Fig. \ref{nK24_massc}), does not show any clear correlation. However, two of the three sources at the bottom of the plot, without an inner disk, have stellar mass close to our lower limit. For this sources inner disk dissipation may be more likely driven by tidal forces or rapid formation of a giant planet rather than photoevaporation, which is expected to be dominant in more massive sources. \citet{luhman08b} noticed that Chamaeleon I, unlike other star forming regions, contains a significant fraction of low-mass stars with inner-disk lifetimes shorter than those of more massive stars.

\begin{figure}[h!]
\centering
\includegraphics[scale=0.5]{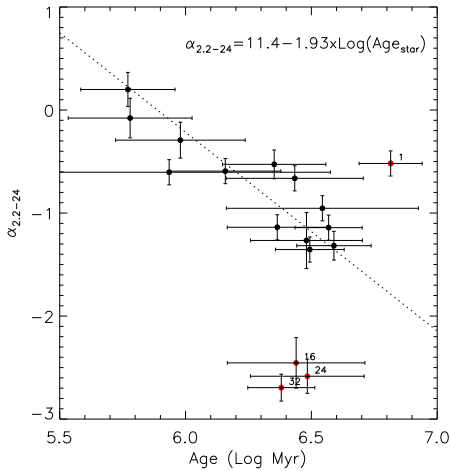}
\caption{Infrared spectral index $\alpha_{2-24}$ as a function of the age. Black dots represent the sources used to fit the main trend (dashed line) and red dots represent the diskless sources.\label{nK24_time}}
\end{figure}
\begin{figure}[h!]
\centering
\includegraphics[scale=0.5]{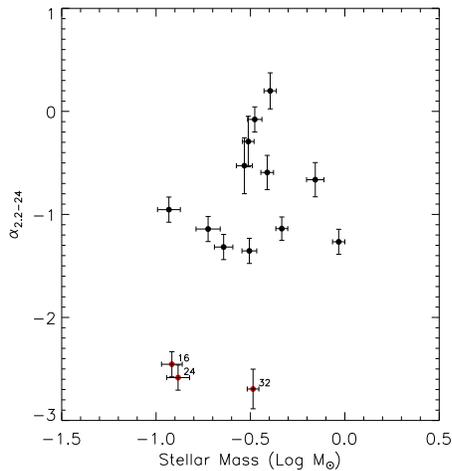}
\caption{Infrared spectral index $\alpha_{2-24}$ in function of the mass. Red dots represent the diskless sources.\label{nK24_massc}}
\end{figure}

In what concerns the mass accretion rates derived from our HST H$\alpha$ photometry, 
several observations have shown a correlation between the mass accretion rate and the age, mass or the IR spectral index of YSOs \citep{muzerolle03, muzerolle05,mohanty05,natta06, alexanderarmitage06, gatti08,sicilia-aguilar10,manara12}. Based on the values reported in Table \ref{s_par} and the magnitudes listed in Table \ref{Tab:phot_known} 
we investigate the correlation between these parameters and the mass accretion rates estimated from our HST images. For sources \#18, \#20 and \#29 we use our new luminosity estimates (Table~\ref{s_par}) to rederive  the mass accretion rates, obtaining $\log \dot{M}_{acc}=-8.2$, -7.9 and -7.38~M$_\odot$/yr$^{-1}$, respectively, and use these values instead of those presented in Table~\ref{Tab:params}.  We also discard sources \# 7 and \#25, as they are confirmed tight binaries \citep[see Section 4.1 and][]{lafreniere08}. 

\begin{figure}[h!]
\centering
\includegraphics[scale=0.5]{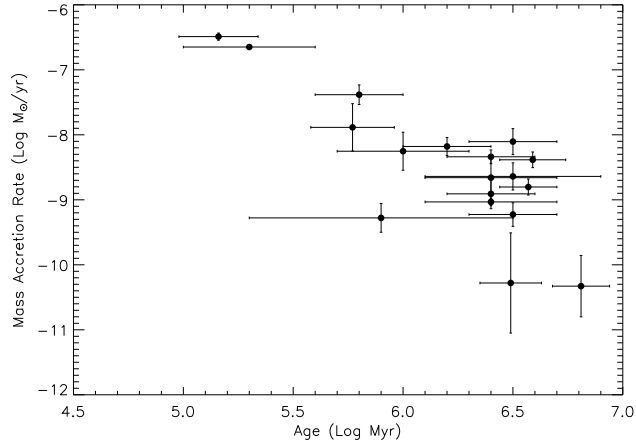}
\caption{Mass accretion rate plotted as a function of the stellar age. The age values are those extrapolated from the SED fitting tool and listed in Table \ref{s_par}. \label{mdot_age}}
\end{figure}

In Fig. \ref{mdot_age} we plot the mass accretion rate as a function of stellar age. The largest accretion rates are found for the two youngest stars, while the majority of older sources show a spread of about 2 orders of mangnitude.  These characteristics,  a general decrease of the mass accretion rate vs. time associated with a large scatter at any given age,  cannot be explained by a any reasonable systematic overestimate of the stellar luminosity (which would make the stars younger while enhancing the estimated accretion luminosity) and are consistent with what typically found in other star forming regions \citep[see e.g.][]{hartmann98}.

\begin{figure}[h!]
\centering
\includegraphics[scale=0.5]{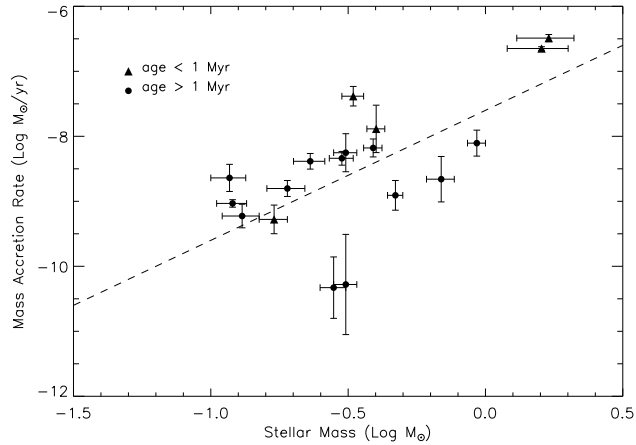} 
\caption{Mass accretion rate plotted as a function of the stellar mass. The mass of the central sources are those extrapolated from the SED fitting tool and listed in Table \ref{s_par}. The dots represent the sources with age$>$1~Myr, whereas the younger sources are indicated by a triangle. The dashed line represent the relation $\dot{M_{acc}}$$\propto$$M_*^2$. \label{mdot_mass}}
\end{figure}

In Fig. \ref{mdot_mass} we show the mass accretion rate as a function of the stellar mass.  The plot shows a gradual rise of the mass accretion rate with the mass of the central source. The dashed line shows the $\dot{M_{acc}}\propto{M_*}^2$ scaling relation reported by various authors \citep{muzerolle03, muzerolle05,mohanty05,natta06} and discussed e.g. in the context of 
accelerated disk clearing by \citet{clarkepringle06} or of systematic differences in disk initial conditions by \citet{alexanderarmitage06}.  

\begin{figure}[h!]
\centering
\includegraphics[scale=0.5]{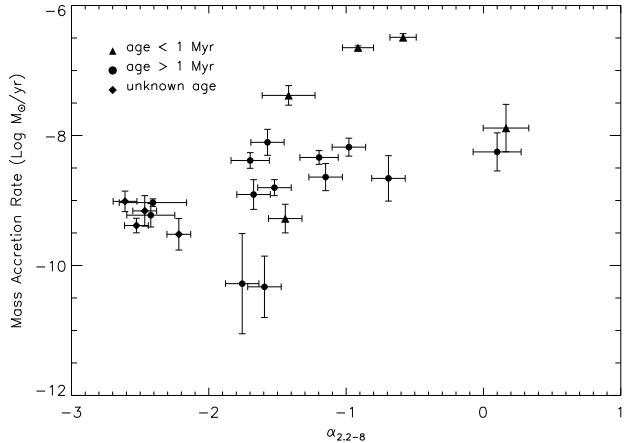} 
\caption{Mass accretion rate plotted as a function of the IR spectral index $\alpha_{2-8}$. \label{mdot_aK8}}
\end{figure}

Finally, in Fig.~\ref{mdot_aK8} we show the relation between the mass accretion rate and the IR spectral index $\alpha_{2-8}$, more sensitive to the warmer dust in the inner disk than the $\alpha_{2-24}$ index. 
The $\alpha_{2-8}$ index is a tracer of both disk evolutionary status and tilt angle. While the scatter of points in 
Fig.~ \ref{mdot_aK8}
may be attributed to the tilt angle, the systematic trend suggests that more evolved inner disks tend to have smaller accretion rates. A similar correlation has been found by 
\citet{sicilia-aguilar10}. 

\section{Conclusions}
We have used the WPC2 instrument onboard the HST to observe 20 fields centered on T Tau stars in the star forming region Chamaeleon~I. Our images, obtained in narrow-filters centered on the [OI], H$\alpha$ and [SII] lines, plus a Johnson-V band equivalent filter, allowed us to detect 31 previously known T Tauri stars. In this paper we have presented the images relative to 10 sources that appear either extended, binaries, or surrounded by a circumstellar disk and/or mass outflow. We have complemented our photometry with a compilation of optical, IR and sub-mm observations, adding new sub-mm data for three sources, together with published values of $T_{eff}$, $A_J$ and $L_{bol}$. Using our H$\alpha$ photometry we have estimated the mass accretion rates for 28 sources. Using all available data, we have reconstructed the optical-IR SEDs of 19 sources and derived a number of disk parameters using the SED-model fitting tool of \citet{robitaille07}. Our main results are:
\begin{enumerate}
\item We resolved 10 Chamaeleon~I sources into binaries or diffuse objects with evidence of circumstellar material, either disks or jets. 
\item The SED fitting shows that most of the sources have mass between 1.7~M$_\sun$ and 0.12~M$_\sun$ and isochronal ages typically ranging between 0.5~Myr and 5~Myr. 
\item The $L_{bol}$ derived from the fitting tool is generally higher than the values reported by \citet{luhman07}, leading to apparently younger ages. The discrepancy increases with the IR spectral index $\alpha_{2-24}$ (see Fig. \ref{nK24_difLuh}). A few sources identified by \citet{luhman04, luhman07} and \citet{luhmanmuench08} as anomalously faint in the near-IR appear to be affected by high dust column density and most probably are associated with disks seen nearly edge-on. 
\item For 13 sources the IR spectral index $\alpha_{2-24}$ appears to decrease with time. The rate of decay would imply that disk dissipation in Chamaeleon~I requires $\simeq10^7$~yr, a relatively long disk lifetime. There are 3 sources that deviate from the main trend, exhibiting an unusual absence of IR excess for their apparently young age. 
\item The mass accretion rate decreases with the stellar age, showing a spread of about two orders of magnitude at $\sim3$~Myr, consistently with what typically found in other star forming regions. The mass accretion increases with the stellar mass roughly following the same scaling relation, $\dot{M}_{acc}\propto M_*^2$ found in other PMS clusters. 
\item
The fact that both the IR spectral index $\alpha_{2-24}$ and the mass accretion rates decrease with our estimated isochronal time suggests that the age spread observed in our sample is real. This is in contrast with the recent suggestion by \citet{jeffries11} that individual stellar ages from the Hertzprung-Russel diagram are unreliable since, at least in the Orion Nebula Cluster, they do not correlate with the presence of disks inferred from near-IR excess. There are clearly several factors that may contribute to the observed luminosity dispersion,  and we have shown that
the case of highly tilted disks is one of those. A comprehensive analysis of the rich phenomenology associated with the presence of accreting circumstellar disks may allow to reveal the intrinsic age spread within a cluster.

\end{enumerate}
{\sl Acknowlegments}: the authors would like to thank Kevin Luhman for discussions and contribution to the source selection process and Basmah Riaz for early analysis of SED fitting. Support for program 11983 was provided by NASA through a grant from the Space Telescope Science Institute, which is operated by the Association of Universities for Research in Astronomy, Inc., under NASA contract NAS 5-26555.

\appendix
\section{SEDs of individual sources}
\begin{description}
\item[2MASS J10533978-7712338 (\#1):] for this source the SED fitting tool is unable to reproduce the relatively flat slope of the Spitzer/IRAC data (3.5-8.0$\mu$m).  
The absence of near-IR excess drives the solution toward a highly tilted system,  i.e. a disk with  81.4$^{\circ}$ inclination with respect to the plane of the sky. The high tilt angle agrees with previous suggestions \citep{luhmanmuench08} that this source (unresolved in our HST images) is probably mainly seen in scattered light. The low bolometric luminosity makes this source the oldest one of our sample. 
\item[ISO-ChaI 10 (\#5):] for this source we obtain a good fit, with stellar mass is in agreement with the value of $\sim$0.18~M$_\sun$ reported by \citet{lafreniere08}. This source has been indicated as a possible binary \citep{lopezmarti04} but remains unresolved in our HST observations. 
\item[Ass Cha T 2-16 (\#9):] also for this source we obtain a a good fit, The stellar mass 0.19~M$_\sun$ is smaller than the previous estimate of 0.26~M$_\sun$ by \cite{lafreniere08}. In Section~\ref{sec:ResolvedSources} we showed some evidence for a spatially resolved PSF, especially in the [OI] line filter.
\item[ESO H$\alpha$-559 (\#10):] for this source we have a sub-mm detection. The best fit indicates a disk seen nearly edge-on (at 87.1$^{\circ}$ tilt) with a disk mass $\sim$2$\cdot$1$0^{-3}$~M$_\sun$ and an age $\sim$7.9~Myr, consistent with \citet{comeron04}. 
\item[V$^*$ UX Cha (\#11):] the SED of this source is typical of a transition disk that has almost entirely dissipated the inner region. The fit returns a stellar mass of $\sim$0.9~M$_\sun$, the same value assigned by \citet{kirkmyers11}.
\item [ISO-ChaI 201 (\#16):] the SED of this source is compatible with a pure photosphere up to 24~$\mu$m. The inner disk has been cleared rather rapidly, as the source seems only $\sim$2.5~My old. This source has been classified as a candidate brown dwarf with spectral type M5.75; our fit assigns a mass of $\sim$0.12~M$_\sun$, close to the lower limit of the grid values. 
\item[Hn 10E (\#17):] the best fit for this source provides a marginally acceptable match to the flat IR SED. The derived stellar parameters are consistent with those reported by \citet{feigelson04}.
\item [CHSM 15991 (\#18):] even if the fitting tool provides a small number of acceptable solutions, the best fit for this source shows  good agreement with the  data. We derive an extreme disk inclination, 87.1$^{\circ}$, in areement with \citet{luhman08b}. Cases like this of extreme disk inclination make the estimates of the absolute stellar lumiosity problematic. \citet{luhman07} estimated for this source $L_{bol}\sim0.0029~$L$_\sun$, which puts below the Main Sequence it in the H-R diagram. Our best fit returns a much higher luminosity, $L_{bol}$$\sim$0.6~L$_\sun$ and an age $\sim$1.0~Myr, for a  $\sim$0.31~M$_\sun$ stellar mass.
\item [CED 112 IRS 4 (\#19):] we obtain a good fit for this source, resolved in our HST images and detected at 870$\mu$m. 
The young age, 0.14~Myr, is well compatible with the HST images showing a young active  source associated with HH 914. The disk inclination, 75.5$^{\circ}$, seems also compatible with the HST morphology. 
\item[ISO-ChaI 225 (\#20):]  for this source we find a small number of acceptable models. In fact, our best fit poorly reproduces the near-IR photometry. The extreme tilt angle, 87.1$^{\circ},$ would imply that the source, unresolved in our HST images, is mainly seen in scattered light. This may well be the case, as \citet{luhman07} estimates $L_{bol}$$\sim$0.013~L$_\sun$ putting the star below the mean sequence. We derive a much higher value, $L_{bol}$$\sim$1.2~L$_\sun$, thus an age of $\sim$0.59~Myr for a mass $\sim$0.40~M$_\sun$. 
\item [V$^*$ WW Cha (\#21):] this source, marginally resolved in our HST images, has two detections in the mm region. SED fitting provides a disk mass of 
$\sim$0.02~M$_\sun$ and an age of $\sim$2$\cdot$1$0^5$~yr.
As for CED 112 IRS 4,  high dust extinction may explain the difference between a stellar mass $\sim$1.6~M$_\sun$ returned from the fit and the estimate of $\sim$0.7~M$_\sun$ by \citet{lafreniere08}. 
\item[ISO-ChaI 232 (\#22):] we obtain a good fit for this source, which shows strong IR excess and mass loss, being associated with objects HH 917 \citep{bally06}, HH 912 and HH 916 \citep{wanghenning06}. The stellar mass returnd by the fitter, 0$.69\pm0.09$~M$_\odot$, is slightly higher than the 0.55~M$_\sun$ estimated by \citet{lafreniere08}. 
\item[Cha T 2-45a (\#23)]: we obtain a good SED fit. The derived stellar mass,  $0.47\pm0.03$~M$_\odot$, is in agreement with the estimate ($\sim$0.51~M$_\sun$) of \citet{lafreniere08}.
\item[Hn 12W (\#24):] The SED shows no evidence of IR excess up to 24$\mu$m. 
 The stellar mass  0$.13\pm0.02$~M$_\odot$ is again in agreement with the estimate ($\sim$0.15~$M_\sun$ ) of \citet{lafreniere08}. 
\item[ISO-ChaI 252 (\#26):] we obtain a genarally good fit except for the IRAC 8micron data point. The stellar mass, $0.12\pm0.02$~M$_\sun$
is in agreement with the estimates of \citet{lafreniere08} and \citet{muzerolle05}, lying out the borderline between stars and brown dwarfs.
\item[HBC 584 (\#27):] for this source with incompled Spitzer/IRAC coverage and strong $\alpha_{2-24}$ index, the best fit indicates a stellar mass $0.39\pm0.03$~M$_\sun$, in agreement with the estimated by \citet{lafreniere08} of 0.35$M_\sun$. 
\item [ESO H$\alpha$-569 (\#29):] a problematic fit for this faint source at optical wavelenghts. The SED  does not show a strong excess at $\lambda<10~\mu$m, but the 24$\mu$m data point is remarkably high.For this source, non detected in the X-rays by \citet{feigelson04}, an extinction $A_k$$\gtrsim$60, possibly due to an edge-on circumstellar disk has been suggested by \citet{luhman07}. The steep rise of the flux at $\lambda<10~\mu$m confirms that the low optical luminosity can be attributed to high dust column density. Our best fit supports this scenario, with a $\sim$87.1$^{\circ}$ disk inclination.  The $L_{bol}$$\sim$0.003~L$_\sun$ derived by \citet{luhman07} places this source below the Main Sequence in the H-R diagram, but integrating the emission reprocessed at longer wavelengths our fit provides $L_{bol}\sim0.1~L_\sun$, with an age $\sim0.63$~Myr and a mass $\sim0.33~M_\sun$.
\item[Hn 21W (\#30):] we obtain a good fit to the SED, with a stellar mass 0$.23\pm0.03$~M$_\odot$ in agreement with the 0.20~$M_\sun$ value reported by \citet{lafreniere08}. 
\item [BYB 53 (\#32):] a class III SED. The age returned by the fit is consistent with the 2~Myr estimated by \citet{gomez03}, indicating that the source has depleted its disk very quickly.
\end{description}


\end{document}